\documentclass[prd,twocolumn,amsmath,amssymb,floatfix,superscriptaddress,nofootinbib]{revtex4}

\usepackage{bm}
\usepackage{amsmath}
\usepackage{epsfig}
\usepackage{color}
\usepackage{natbib}
\usepackage{graphicx} 

\makeatletter
\renewcommand\footnoterule{%
  \vspace{-5pt}
  \kern-3\p@\hrule\@width.4\columnwidth%
  \kern10\p@}
\makeatother

\def\be{\begin{equation}}
\def\ee{\end{equation}}
\def\ba{\begin{eqnarray}}
\def\ea{\end{eqnarray}}

\newcommand{\Npiv}{N_{\rm pivot}}

\newcommand{\aend}{a_{\rm end}}
\newcommand{\aeq}{a_{\rm eq}}
\newcommand{\arh}{a_{\rm th}}
\newcommand{\wprim}{w_{\mathrm{ prim}}}

\newcommand{\rhorh}{\rho_{\rm th}}
\newcommand{\rhoeq}{\rho_{\rm eq}}
\newcommand{\rhoend}{\rho_{\rm end}}

\newcommand{\vend}{V_{\rm end}}
\newcommand{\Mpl}{M_{\rm Pl}}

\definecolor{darkgreen}{cmyk}{0.85,0.2,1.00,0.2} 
\definecolor{purple}{cmyk}{0.5,1.0,0,0}

\newcommand{\ModeCode}{{\sc Mode\-Code}}
\newcommand{\MultiNest}{{\sc MultiNest}}
\newcommand{\CAMB}{{\sc CAMB}}
\newcommand{\CosmoMC}{{\sc CosmoMC}}
\newcommand{\codename}{{\ModeCode}}

\begin{document}
\title{Bayesian Analysis of Inflation II: Model Selection and Constraints on Reheating}

\author{Richard Easther}\email{r.easther@auckland.ac.nz}
\affiliation{Department of Physics, Yale University, New Haven, CT 06520, U.S.A.}
\affiliation{Department of Physics, University of Auckland, Private Bag 92019, Auckland,
New Zealand}

\author{Hiranya V. Peiris}\email{h.peiris@ucl.ac.uk}
\affiliation{Department of Physics and Astronomy, University College London, London WC1E 6BT, U.K. \\ \mbox{}}

\date{\today}

\begin{abstract}
\baselineskip 11pt
We discuss the model selection problem for inflationary cosmology.  We couple \ModeCode, a publicly-available numerical solver for the primordial perturbation spectra, to the nested sampler \MultiNest, in order to efficiently compute Bayesian evidence.  Particular attention is paid to the specification of physically realistic priors, including the parametrization of the post-inflationary expansion and associated thermalization scale. It is confirmed that while present-day data tightly constrains the properties of the power spectrum, it cannot usefully distinguish between the members of a large class of simple inflationary models.  We also compute evidence using a simulated Planck likelihood, showing that while Planck will  have more power than WMAP to discriminate between  inflationary models, it will not definitively address the inflationary model selection problem on its own. However, Planck will place very tight constraints on any model with more than one observationally-distinct inflationary regime --  e.g. the large- and small-field limits of the hilltop inflation model -- and put useful limits on different reheating scenarios for a given model. 
\end{abstract}

\maketitle

\section{Introduction} \label{sec:intro}

The prospect that astrophysical observations   probe  the properties of the very early universe and test physics at  energies approaching the Planck scale is one of the most exciting aspects of modern cosmology. Inflation  \cite{Guth:1980zm} provides a concrete realization of this possibility,  as the primordial perturbations are generated during a well-defined epoch in the early universe (in almost all versions of the scenario) at energies far above the TeV scale.  

A common approach to constraining inflationary models is to project their predictions for the scalar spectral index and tensor amplitude  onto the corresponding  likelihood contours (see e.g. Refs \cite{Spergel:2006hy,Komatsu:2008hk,Komatsu:2010fb}).  This simple and largely qualitative  method is usually sufficient with present-day data, but has a  number of  shortcomings.  In particular, it is not easily extended to scenarios with several non-trivial observable parameters, e.g. a running spectral index or features in the power spectrum. Likewise, the predictions of inflationary models with two or more free parameters (e.g. natural inflation) can  correspond to domains of the  $(n_s,r)$ plane with nontrivial shapes,  rendering interpretation difficult.    Most importantly, this approach does not address the model selection problem for inflation in a rigorous, quantitative way.
 
Identifying optimal methods for constraining  inflationary models is a  long-standing problem, and a large number of schemes  have been explored    (e.g. Refs \cite{Copeland:1993ie, Copeland:1993jj,Lidsey:1995np,Hansen:2001eu,Leach:2002ar, Leach:2002dw, Leach:2003us,Kinney:2006qm,Peiris:2006ug,Peiris:2006sj,Martin:2006rs,Lesgourgues:2007gp, Lesgourgues:2007aa, Peiris:2008be,Adshead:2008vn,Hamann:2008pb,Powell:2008bi,Kinney:2008wy,Martin:2010hh}).  Anticipated developments in observational cosmology -- particularly the Planck cosmic microwave background (CMB) survey \cite{:2006uk} and forthcoming large scale structure surveys -- make it crucial to definitively address this topic. This is the second in a sequence of papers (begun with Ref.~\cite{Mortonson:2010er}, henceforth referred  to as Paper~I) that  reviews and extends previous work, and develops \ModeCode\footnote{http://zuserver2.star.ucl.ac.uk/$\sim$hiranya/ModeCode/ModeCode}, a publicly available and well-tested suite of computational tools for constraining inflationary models. In Paper I we estimated the free parameters of individual inflationary models, coupling a numerical solver for the primordial perturbations to the cosmological Monte Carlo Markov Chain (MCMC) code \CosmoMC\ \cite{Lewis:2002ah} and the standard Boltzmann code \CAMB\ \cite{Lewis:1999bs}.

In this paper, we turn to the issue of model selection.  This problem is distinct from parameter estimation, although the tools used here tackle both  tasks simultaneously. Frequentist approaches to model comparison (e.g. $P$-values, $\Delta \chi^2$ and $N\sigma$ deviations from the mean)  have conceptual and practical shortcomings when applied to the properties of the universe as a whole, given that they are  defined relative to a large and usually unspecified ensemble (see e.g. Ref. \cite{Cox:1946}).  Conversely, Bayesian methods naturally rise to the challenges presented by cosmological model selection.  

In Bayesian terminology, an inflationary scenario defines the model prior, which in turn defines the parameter volume of a given model. The prior includes both the usual ``post-inflationary'' cosmological parameters,  plus the free parameters within the inflationary sector itself.   We will see that a complete and self-consistent specification of the prior is crucial to successfully addressing the model selection problem for inflation, while the parameter estimation process is typically insensitive to many of these choices.

The Bayesian model selection statistic, often called the {\em evidence\/} $E$, is the ``model-averaged likelihood'', i.e. the integral of the likelihood $\mathcal{L}$ over the parameter volume $\{\alpha_1,\cdots,\alpha_M\}$, 
\begin{equation} \label{eq:Evidencefull}
E=\int d\alpha^M P(\alpha_i)\mathcal{L}(\alpha_i)\, ,
\end{equation}
weighted by the prior $P(\alpha_i)$, which is normalized so that $\int  d\alpha^MP(\alpha_i) \equiv 1$.  For  uniform priors  $P(\alpha_i)$ is a constant (or zero),  and the weighting is the inverse parameter volume, or
\begin{equation}\label{eq:flatprior}
E=\frac{1}{\text{Vol}_M}\int d\alpha^M \mathcal{L}(\alpha_i)\, .
\end{equation}
This is a multidimensional integral over a  computationally expensive  integrand. Consequently, in what follows we pay attention to the  practical challenges associated with computing $E$  for inflationary models.  
 
The evidence is not an absolute quantity: the ratio of the evidence values for two models expresses the relative ``betting odds'' that  these  models are  responsible for the observed state of the universe.    A model is a combination of both a physical parameterization of the early universe and a set of permitted ranges (priors) for each of the free parameters. The model specification can include a relative preference for some parameter values over others, codified by $P(\alpha_i)$.   It can also include an overall preference for one model over another.

Qualitatively, the role of Bayesian evidence in model selection can be understood as follows. Restricting our discussion to uniform priors, if a model $M$ predicts a ``wide'' parameter range,  the allowed $\alpha_i$ enclose a volume much larger than the peak of $\mathcal{L}$. Consequently,  the evidence $E$ computed for $M$ is reduced, relative to $E'$, computed for an otherwise identical model $M'$ where the allowed parameter ranges  contain  only the region in which $\mathcal{L}$ is appreciably different from zero. Physically, the predictions of the model $M$  are less tightly specified than those of $M'$, and the tight match between prediction and observation for $M'$ is ``rewarded'' by the evidence criterion. This also underlines the importance of setting the priors independently of the data, since choosing $M'$ {\em a posteriori\/} is not a prediction. Conversely, a  model $M''$ for which   $\mathcal{L}$  is small everywhere in  $\alpha_{M''}$ will have $E''\ll E'$ and $E'' \ll E$, reflecting the inability of $M''$ to fit the data for any choice of parameters. 

 Evidence ratios are often compared  using the Jeffreys scale\footnote{ It is important to note that while the Bayesian approach described above is the only self-consistent framework for model selection \cite{Cox:1946}, the Jefferys scale is an ad hoc set of preferences for where to draw the boundaries -- individual scientists can be more or less conservative in the betting odds at which they will prefer one model over another.} \cite{Jeffreys1998BK},  which rates $\Delta\ln E<1$  as being ``not worth a bare mention'', whereas $\Delta\ln E > 5$  is regarded as ``highly significant''.  In what follows, we pay careful attention to the specification of the  range of priors associated with a given inflationary model.  

We proceed by adapting \ModeCode\ to use the nested sampler, \MultiNest\  \cite{Feroz:2007kg,Feroz:2008xx}  --  an alternative  to  the default MCMC  exploration of the parameter space in \CosmoMC.  \ModeCode\ is based on an algorithm developed in Ref.~\cite{Adams:2001vc}; the underlying code was also used in Refs~\cite{Peiris:2003ff,Mortonson:2009qv}.    \MultiNest\ rapidly computes Bayesian evidence while simultaneously providing  parameter estimates.  In addition, nested sampling is often much more efficient than MCMC sampling when the  likelihood surface is significantly non-Gaussian -- a situation encountered with even relatively simple inflationary models.\footnote{Nested sampling also efficiently explores scenarios where the likelihood contours have a nontrivial topology, which can occur in inflationary models  with ``features'' or steps (e.g. Ref. \cite{Mortonson:2009qv}).}  

The problem of model selection in early universe cosmology is not a new one, and has been tackled by a number of  authors (e.g. Refs \cite{Liddle:2006tc,Parkinson:2006ku,Kawasaki:2009yn,Kilbinger:2009by,Martin:2010hh}).  In particular, the present analysis overlaps with that of Ref. \cite{Martin:2010hh}, which also computed evidence by coupling \MultiNest\ to a numerical solver for the inflationary perturbation equations. The specific contributions made by the present work include a careful discussion of the specification of both the inflationary models and associated model priors, and the interplay between the choice of prior and the computational efficiency of the code.  We also analyze inflationary models (hilltop and natural inflation) which  occupy non-trivial domains in the $(n_s,r)$ plane, and explore the impact of post-inflationary physics on the model selection problem. Finally, as part of the ``warm-up'' for Planck, we analyze models using both a simulated Planck likelihood and current Wilkinson Microwave Anisotropy Probe (WMAP) data  \cite{Komatsu:2010fb}.
   
The paper is structured as follows: Section II provides a very brief review of inflationary phenomenology and its implementation within \ModeCode, and the parameterization of the post-inflationary behavior of the universe. We discuss the specification of priors for  inflationary scenarios in Section III, and present evidence ratios computed with \ModeCode\ for a sample of inflationary models, for both current WMAP data and a Planck simulation in Section~IV.

This paper is the second in a sequence of articles that  survey, develop and implement theoretically optimal and computationally efficient methods for applying astrophysical constraints to the inflationary  phase.  \ModeCode\ is thus a work in progress, but regularly updated versions of the code are available for download.

\section{ModeCode and MultiNest: Numerical Implementation}
 
Simple models of inflation are governed by the Friedman equation and sourced by a minimally coupled scalar field $\phi$.  As reviewed in Paper I,  the inflaton obeys the  Klein-Gordon equation in an expanding background
\begin{equation}
\label{eq:phieom} \ddot{\phi} + 3 H \dot{\phi} + \frac{dV(\phi)}{d \phi} = 0 \, .
\end{equation} 
As usual $V(\phi)$ is the potential and specifies the  model, while  $H$ is the Hubble parameter and dots denote derivatives with respect to  time.   This formalism  implicitly assumes that inflation lasts long enough to  erase any primordial spatial curvature and other ``pre-inflationary'' relics.     We specify dimensionful quantities relative to the reduced Planck mass,  $\Mpl = 2.43 \times 10^{18}$~GeV.   \codename\   computes the  power spectrum by numerically solving the evolution equations for the scalar and tensor perturbations, as described in Paper I and Ref. \cite{Adams:2001vc}.  

Comoving scales in the primordial universe are connected to present-day astrophysical scales by the matching equation \cite{Liddle:2003as,Dodelson:2003vq,Alabidi:2005qi}.   The observed power spectrum is a function of the rate at which modes reenter the horizon after inflation, which is in turn fixed by the effective equation of state during this era.   Consequently, as explained in Refs \cite{Adshead:2010mc,Mortonson:2010er}, observational tests of inflationary models simultaneously explore both the inflationary phase  and the subsequent expansion history.  

For representative models of inflation, Planck  can be expected to constrain the number of $e$-folds $N$ to within $\Delta N \sim \pm 3$ at the 68\% confidence level \cite{Adshead:2010mc}. Consequently, we wish to work with a version of the matching equation which explicitly accounts for the relativistic neutrino contribution to the universe (which delays matter-radiation equality), and the dark-energy driven accelerated expansion of the present-day universe.         \codename\ relates the primordial perturbation spectrum to present day scales via
\begin{eqnarray}
 \aend &=& \frac{\aend}{\arh} \frac{\arh}{\aeq} \frac{\aeq}{a_0} a_0 \nonumber \\
    &=& \left( \frac{\rhorh}{\frac{3}{2}\vend}\right)^\frac{1-3\wprim}{12(1+\wprim)}  \left( \frac{\rhoeq}{\frac{3}{2}\vend}\right)^{\frac{1}{4}} \frac{\aeq}{a_0} a_0 \, , \label{eq:matchina}
\end{eqnarray}	
where $a(t)$ is the usual scale factor. Recall that inflation ends when $\ddot{a} =0$ which, for a scalar field, occurs when $V=\vend = \frac{2}{3} \rhoend$, where $\rho$ is the energy density.  Following Refs \cite{Adshead:2010mc,Mortonson:2010er}, we assume that the universe is definitely thermalized when $\rho = \rhorh$ and $a=\arh$, and has a conventional thermal history thereafter.  Note we are not stipulating that   $\rhorh$ is the  density at which the universe {\em becomes\/} thermalized, merely that the universe is  thermalized as the density passes through this value. The eq subscript refers to values at matter-radiation equality.

Equation~(\ref{eq:matchina})   assumes that the primordial universe is dominated by a perfect fluid with density and pressure related by $p=\wprim \rho$ at  densities $\rho>\rhorh$, where $\wprim$ is the  effective equation of state during this phase.\footnote{The detailed equation of state during the primordial dark age is needed to compute the amplitude of stochastic gravitational wave backgrounds at scales inside the horizon while $\rho > \rhorh$  \cite{Boyle:2005se}. However, for the foreseeable future strong constraints from stochastic gravitational wave backgrounds are unlikely to be a practical concern.}

In contrast to Paper I, \ModeCode\  now includes the minimal thermalization scale $\rhorh$ as part of the model specification.  The distinction between  $\rhorh$ and the density at the {\em onset\/}  of thermalization   is of crucial importance. Recall that while the ``hot big bang'' is the basis of  modern cosmology, there is no firm evidence that the universe was thermalized prior to neutrino decoupling and the onset of nucleosynthesis. The period between these two epochs -- which spans a range of $\sim10^{18}$  in energy -- is the so-called {\em primordial dark age\/} \cite{Boyle:2005se}.    

There is no evidence that the universe is {\em not\/} thermalized during the post-inflationary epoch, but equally we have no evidence that it is radiation dominated during this phase. The relevance to this discussion is that the matching between inflationary and astrophysical scales is a function of the rate at which modes re-enter the horizon, and this is determined by the expansion rate of the universe, which is in turn fixed by the effective equation of state.  For instance, the primordial universe can include  phases of coherent oscillations \cite{Martin:2010kz,Easther:2010mr}, resonance \cite{Traschen:1990sw,Allahverdi:2010xz}, kination \cite{Spokoiny:1993kt,Chung:2007vz}, secondary or thermal inflation \cite{Lyth:1995ka}, moduli domination \cite{Banks:1993en,deCarlos:1993jw},  primordial black hole domination \cite{Anantua:2008am}, or a frustrated cosmic string network \cite{Burgess:2005sb}, all of which lead to an expansion rate that differs from that of a radiation dominated universe.   The effective equation of state $\wprim$ is thus an appropriately weighted average of the instantaneous equation of state during  the primordial dark age.

 Physically, the fundamental  parameter that sets the observable perturbation spectrum is the value of $\phi$ at which the pivot mode leaves the horizon. However, $\phi$ can both increase and decrease with time (depending on the shape of the potential);  its absolute value can be rescaled by a shift $\phi\rightarrow \phi + \phi_0$; and the overall range of $\phi$ during inflation varies greatly between models. Consequently, we treat the remaining number of $e$-folds $N$ after the pivot scale leaves the horizon as a free parameter, since this quantity has a consistent interpretation across  models.   
 
 The impact of the primordial dark age on  large scale structure and CMB data is thus captured by  $\wprim$ and $N$.    \ModeCode\ draws a value of $N$, then computes $\wprim$, which is a derived parameter -- models for which $\wprim$ lies outside a specified range are excluded from the prior.   The  value of $\rhorh$ is  specified as part of the prior, as we discuss in the following Section - it can be fixed by both data-driven constraints and theoretical inferences about the thermalization scale. 
 
  By definition,  at matter-radiation equality, $2 \rho_{\rm rad} = \rho$.   All three neutrino species are relativistic at this transition  \cite{Komatsu:2010fb}, and we have the well-known result 
\begin{equation}
\rho_{\rm rad}  = \frac{\pi^2}{15} \left[ 1 + \frac{7}{8} N_\nu \left( \frac{4}{11} \right)^{4/3} \right] T_{\rm CMB}^4 \, .
\end{equation}
Since $T_{\rm CMB}(t) a(t)$  is constant through neutrino freeze-out and the onset of  dark energy domination, we obtain
\begin{eqnarray}
 \ln(\aend) &=&   \frac{1-3\wprim}{12(1+\wprim)} \ln{\left(\frac{\rhorh}{\rhoend} \right)}-\frac{1}{4}\ln{\left( \frac{\rhoend}{\Mpl^4}\right) } + \nonumber \\
    && \frac{1}{4} \ln{\left[1 + \frac{7}{8} N_\nu \left(\frac{4}{11}\right)^{\frac{4}{3}}\right]} -71.32 \label{eq:match}
\end{eqnarray}	
using the observational result $T_{{\rm CMB},0} =  2.348 \times  10^{-4}$~eV.  \codename\ can marginalize over $N$, the remaining number of $e$-folds of inflation as a specified pivot scale leaves the horizon, and $N$ effectively scales with $\ln{\aend}$.  The effective number of neutrino species $N_\nu$  only has a small impact on $\ln{\aend}$, and therefore we do not simultaneously estimate neutrino number alongside inflationary parameters. Consequently, \codename\ fixes $N_\nu= 3.04$, moving the constant in equation~(\ref{eq:match}) to $-71.21$.    At this point, given the  specified value of $\rhorh$,  equation~(\ref{eq:match}) can be inverted to yield $\wprim$.
 
\section{Inflationary Priors and the Computation of Bayesian evidence}

\subsection{Inflationary Models}
 
We consider the same set of models as Paper I. These include single parameter models based on the potential
\begin{equation}\label{eq:singleterm}
 V = \lambda  \frac{\phi^{n}}{n} 
\end{equation}
for fixed $n$, axion-motivated ``natural inflation'' \cite{Freese:1990rb} with
\begin{equation}
V(\phi) = \Lambda^4 \left[1+\cos\left(\frac{\phi}{f}\right)\right] \, ,
\label{eq:natural}
\end{equation}
and ``hilltop inflation''  \cite{Kinney:1995cc,Kinney:1998md,Easther:2006qu} with
 \begin{equation}
V(\phi) = \Lambda^4 - \frac{\lambda}{4}\phi^4 \ ,
\label{eq:hilltop}
\end{equation}
for which the tensor-scalar ratio  can be arbitrarily small.

\begin{table}
\begin{center}
\begin{tabular}{|l|c|}
\hline
\hline
Model & Priors   \\
\hline 
\hline
$n=2/3$ & $-13<\log_{10}\lambda<-7$  \\ \hline
$n=1$ & $-13<\log_{10}\lambda<-7$  \\ \hline
$n=2$ & $-13.5<\log_{10} m^2<-8$  \\ \hline
$n=4$ & $-16<\log_{10}\lambda<-10$  \\ \hline
Natural & $-5<\log_{10}\Lambda<0$ \\
 & $0<\log_{10} f<2.5$ \\ \hline
Hilltop & $-8<\log_{10}\Lambda<-1$    \\
 & $-17<\log_{10}\lambda<-10$   \\
\hline
\hline
\end{tabular}
\end{center}

\begin{center}
\begin{tabular}{|l|c|}
\hline
\hline
Matching   & Prior \\
\hline 
\hline
$\wprim$ & $-1/3\le \wprim \le 1$  \\ \hline
$N$ &   $20\le N \le 90$  \\
\hline
\hline
\end{tabular}
\end{center}

\begin{center}
\begin{tabular}{|l|c|}
\hline
\hline
Variable &   Prior   \\
\hline 
\hline
Baryon fraction  & $0.015 < \Omega_b h^2 < 0.035$ \\
 \hline
Dark matter  &  $0.05 < \Omega_{dm} h^2 < 0.2$  \\
\hline
Reionization & $0.01 < \tau <  0.25$ \\
\hline
Projected acoustic scale & $0.8 <\theta < 1.2 $ \\
\hline
Sunyaev-Zel'dovich Amplitude &  $0<A_{\rm SZ} < 2$ \\
\hline
\hline
\end{tabular}
\end{center}

\caption{We list the inflationary models (values of $n$ refer to specific cases of  equation~(\ref{eq:singleterm})) and the non-inflationary free parameters, and the priors for each parameter.  We assume that $\Omega_k$ is zero in all cases, and all parameters are drawn from a uniform prior. The matching parameters allow us to  marginalize over the expansion history during the primordial dark age,  connecting the inflationary era to astrophysical scales. We also apply a cut on the amplitude of the primordial power spectrum, as described in the text. The pivot scale used to compute $N$ is $k_\star = 0.05$~Mpc$^{-1}$.  }
\label{tab:priors}
\end{table}

\begin{figure}[t]
\centerline{\epsfig{file=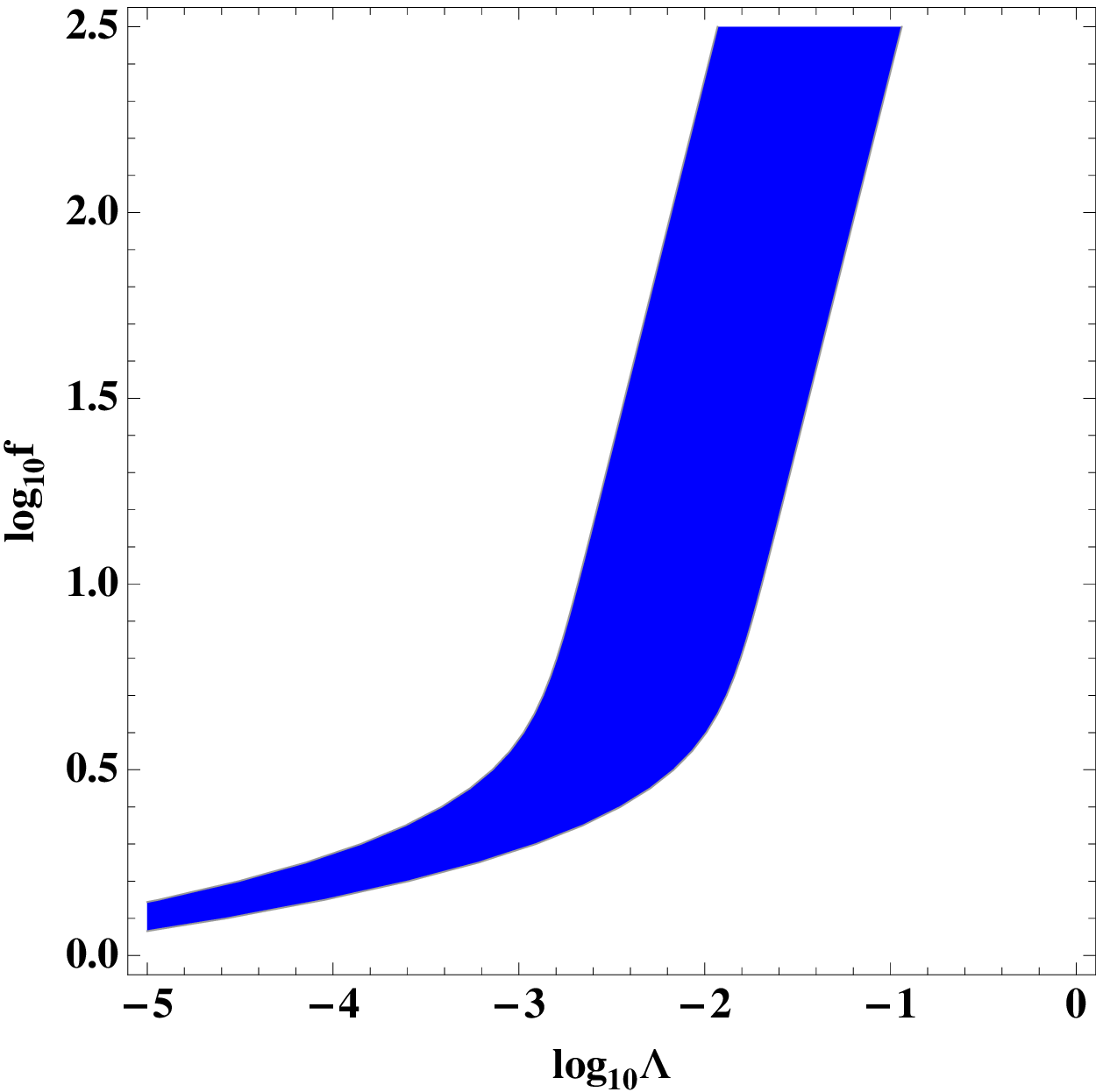, width=3in}}

\mbox{}

\centerline{\epsfig{file=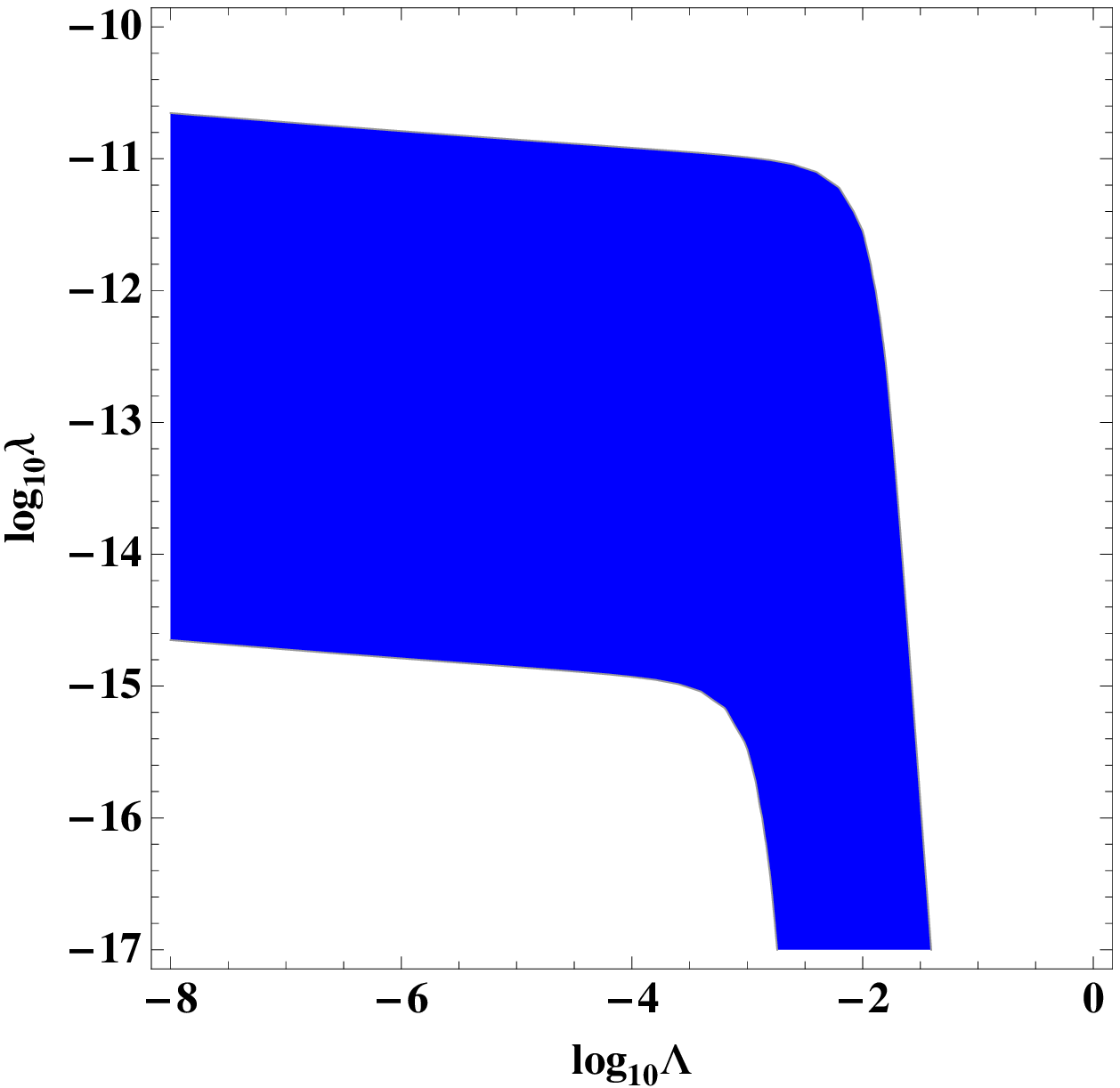, width=3in}}

\caption{Parameter ranges for natural (top) and hilltop (bottom) models.  The shaded regions shows parameter values for which the spectral amplitude $10^{-11} \le A_s \le 10^{-7}$ respectively. Points outside these regions are excluded from the prior.  \label{fig:priorAs} }

\end{figure}

\subsection{Parameters and Priors}

Specifying physically realistic and self-consistent priors is a key prerequisite to making meaningful statements about the relative merits of competing inflationary models.   In Bayesian inference, the prior encodes any ``background knowledge'' or physical assumptions that contribute to the construction of the model.  Crucially, the prior must be specified independently of the data  used to construct the likelihood $\mathcal{L}$.  This is a particularly acute problem for inflationary model selection, since analyses of specific models of the early universe typically rely on large datasets with well-known properties (e.g. the WMAP results), making it impossible to perform a genuinely blind analysis. 

A  prior consists of both a physical model and the allowed ranges of its  free parameters, specified by $P(\alpha_i)$.  Choosing these ranges raises several subtle problems.    For example, in $\Lambda$CDM cosmology, the primordial power spectrum is specified  by the spectral index $n_s$, amplitude $A_s$ and  (arbitrary) pivot $k_\star$, as a function of comoving wavenumber, $k$
\begin{equation}
P(k) = A_s \left(\frac{k}{k_\star} \right)^{n_s-1} \, .  \label{eq:pk}
\end{equation}
At the time of writing the default range for $n_s$ within \CosmoMC\  is $0.5< n_s<1.5$.    This is a uniform prior, so this choice stipulates that  $n_s =0.5$ is as likely as $n_s=1.0$. This  proposition is wildly at odds with even mild observational constraints on the primordial spectrum.\footnote{For example, upper limits on the abundance of primordial black holes rule out the upper end of this range \cite{Green:1997sz} while large scale structure constraints rule out the lower end of this range.}  For MCMC-based parameter estimation this is  a purely formal issue:  parameter ranges have no impact on the  results, so long as regions with nontrivial $\mathcal{L}$ are not excluded. MCMC samplers naturally focus on the region(s) of highest likelihood, and computational performance is governed by  the proposal distribution rather than the total prior volume, and there is no cost to specifying very wide priors.

Conversely,  evidence probes the whole parameter space, and always depends on the  ranges specified in the prior.  In many cases -- including $n_s$ in the example above -- there is no unambiguous choice for the parameter range,  effectively inducing a ``theory error'' in the computed evidence.  For instance, choosing $0< n_s<2$ or $0.8 < n_s < 1.2$  changes  $\Delta  \ln E$ for $\Lambda$CDM by $\sim \pm1$, but there is no obvious basis for any of these specific choices \cite{Easther:2011wh}.  For very large $\Delta  \ln E$ this will not significantly change the outcome of model selection.  However, in the next Section we find that Planck  typically yields $\Delta  \ln E \sim 2$  when evidence is computed for simple inflationary models. Consequently, this  ambiguity must be tightly controlled.

It may appear that inflationary potentials are less arbitrary than  equation~(\ref{eq:pk}), a purely empirical characterization of the spectrum.  However, very few inflationary potentials are derived from  a well-controlled fundamental theory, and in many cases the potential is simply written down by the model-builder.    Consequently, we have little {\em a priori\/} information on the likely values of their free parameters\footnote{One of the few well-motivated constraints on the inflationary parameter space is the Lyth bound \cite{Lyth:1996im}.  Naively applied, this  rules out the entire parameter range over which $\mathcal{L}$ is non-zero for models with $V(\phi)\sim\phi^n$, by excluding parameter choices with a trans-Planckian field-excursion.  However,  the Lyth bound is based on an effective field theory argument, and explicit counter-examples to this generic prohibition are well known   \cite{Silverstein:2008sg,McAllister:2008hb,Dong:2010in}.  Whether the Lyth bound is incorporated in the  prior thus reflects the judgement of the modeler, as is standard in a Bayesian analysis.}   and our expectations for their values are largely determined from the properties of the predicted power spectrum, rather than fundamental physical considerations.

The strongest empirical constraint on the inflationary parameter space is the amplitude of the primordial power spectrum.  This  is a free parameter in most  models, and successful structure formation in a universe dominated by cold dark matter  has long been known to require  primordial  fluctuations  with $\delta \rho / \rho \sim 10^{-5}$, or $A_s\sim 10^{-10}$  (see e.g. Refs \cite{Zeldovich:1972zz,Linde1990Bk}).  Thus we can immediately reject models for which  $A_s$ is far from this value, without reference to the data used to construct $\mathcal{L}$.   Consequently, in this analysis, regions of parameter space which do not yield $10^{-11} \le A_s \le 10^{-7}$ are excluded via the prior.    This range is generous relative to current estimates of $A_s$ (e.g. Ref. \cite{Komatsu:2010fb}). However, our results do not  depend strongly on the overall range, and we wish to work with constraints that reflect  only {\em a priori\/} knowledge that is genuinely independent of recent high-precision astrophysical data.  

For the single parameter models this requirement  defines the  range of $\lambda$ in equation~(\ref{eq:singleterm}). For these cases our approach is equivalent to that of Ref. \cite{Martin:2010hh}, in which a prior is specified for $\ln(A_s)$ and the corresponding inflationary variable is then a derived parameter.  However, for generic multi-parameter models  an $A_s$-based cut may select a nontrival region of parameter space, as  happens for the two cases considered here (Figure~\ref{fig:priorAs}).  Without the $A_s$-based cut in the prior, the parameter volume for both natural and hilltop inflation would be rectangular, and the corresponding evidence values computed for these models would be lowered accordingly.   Conversely, for these models  eliminating one parameter by fixing $A_s$  would induce a strongly  non-uniform prior on the remaining parameter(s), potentially biasing both the computed evidence and estimated parameter values.

Generally, the free parameters in the inflationary potentials can take a large range of values corresponding to unknown scales in high energy particle physics. Consequently, it is appropriate to use logarithmic priors on these parameters, or equivalently, to impose uniform priors on their logs. 

As noted previously, the thermalization scale is only weakly constrained by direct cosmological measurements --  nucleosynthesis and the cosmological neutrino background require  $\rhorh^{1/4} \gtrsim \mathcal{O}(100)$ MeV,  far below the inflationary scale in almost all models.   However, the  thermalization mechanism must produce Standard Model particles from  the inflaton or its immediate decay products which, in almost all scenarios, are not part of the Standard Model.  For this to occur, at least some Standard Model particles must necessarily couple to fields outside the Standard Model. The resulting interactions generically lead to loop corrections to  precision electroweak observables, which must be small in order to avoid conflicts with existing experiments.  The easiest way to guarantee that these couplings are not problematic is for reheating to occur at energies far above the electroweak scale.  It is possible to construct models in which reheating occurs at sub-TeV scales, but these scenarios are tightly constrained -- so the ``natural'' expectation is  that the universe is thermalized at temperatures above $1$~TeV.

Setting $\rhorh^{1/4} \sim \mathcal{O}(100)$ MeV with a flat prior on $N$ stipulates that it is just as likely that the universe thermalizes at (say) $1$~GeV as  at $10^{10}$  GeV.  Similarly to the situation with the ``cosmological'' parameters, this choice of prior therefore amounts to a statement that is at odds with our understanding of fundamental physics. Rather than quantify the relative likelihood of thermalization at different scales, we checked that the computed values of the evidence are not sensitive to the specific choice of $\rhorh$.  We also run the sampler with an instant reheating prior, specifying that thermalization occurs immediately after the end of inflation. In this case $\wprim=\frac{1}{3}$ by definition, rendering $N$  a derived parameter.\footnote{This approach is again different from that of Ref.  \cite{Martin:2010hh}, where $N$ is assigned a uniform prior, and a lower bound that corresponds to an MeV scale reheating temperature.}

\ModeCode\  allows the user to specify minimal and maximal values of $\wprim$ --  we expect  $\wprim \le1$ to avoid a superluminal sound speed and $\wprim \ge -1/3$, so  $\ddot{a} \le 0$, and the post-inflationary universe is not undergoing accelerated expansion.  Given that the predictions of inflationary models are sensitive to the details of the post-inflationary expansion,  scenarios which make a specific prediction for the physics of this epoch should be evaluated with an appropriate (and tighter) prior for  $\wprim$.   

From a  practical perspective, if the likelihood is effectively zero over a large fraction of the  parameter volume, parameter estimation with  \MultiNest\  is far more computationally intensive than with the  MCMC sampler in \CosmoMC,  often  failing to converge in any reasonable amount of time.  With  ``realistic'' priors the two codes  have similar runtimes (to within factors of a few) for single field models, while  \MultiNest\  is at least an order of magnitude faster than an MCMC sampler for natural or hilltop inflation models, which have nontrivial likelihood contours.    Our full set of priors -- for both the inflationary physics and regular cosmological parameters -- is given  in Table~\ref{tab:priors}.    We typically run \MultiNest\ with 800 live points and  tolerance and efficiency parameters set to $0.3$. With non-trivial likelihood topologies or very thin likelihood contours in some dimensions, we increase the number of  live points by a factor of a few.

\begin{figure*}[tb]
\centerline{\epsfig{file=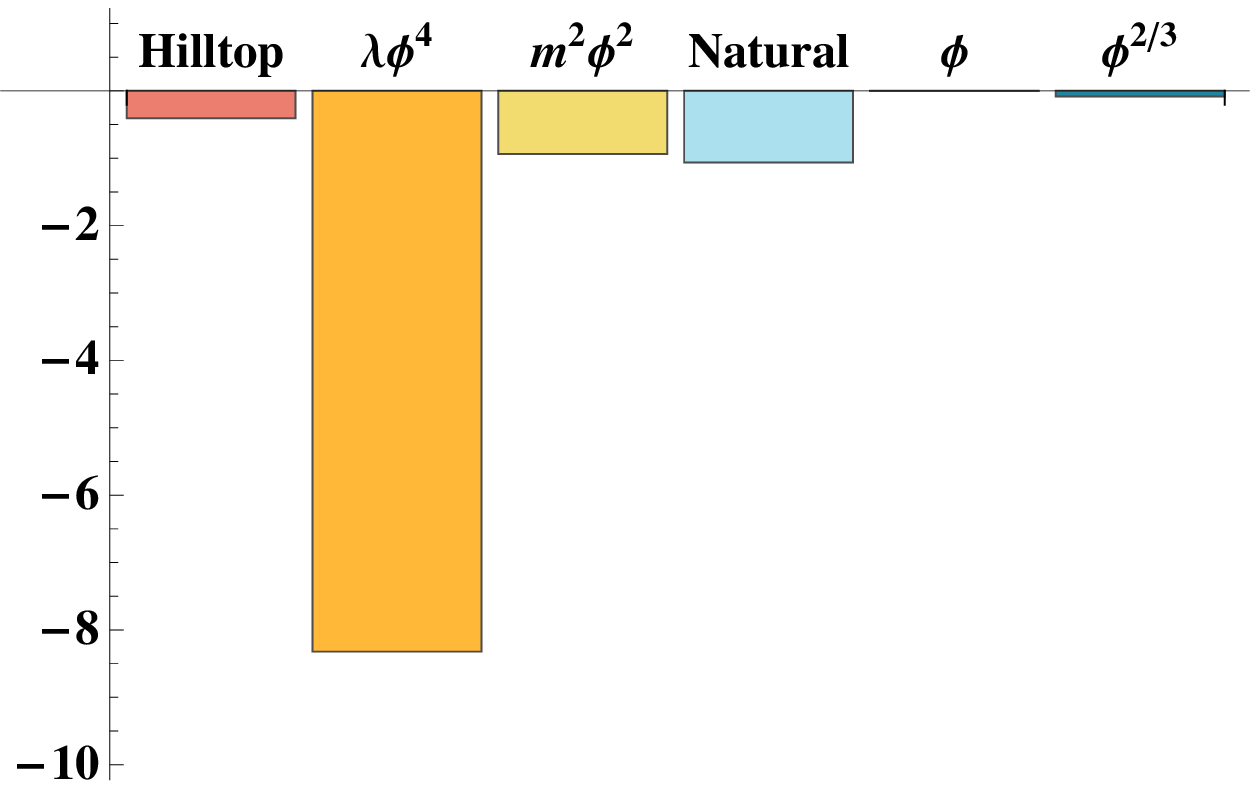, width=3in}  \epsfig{file=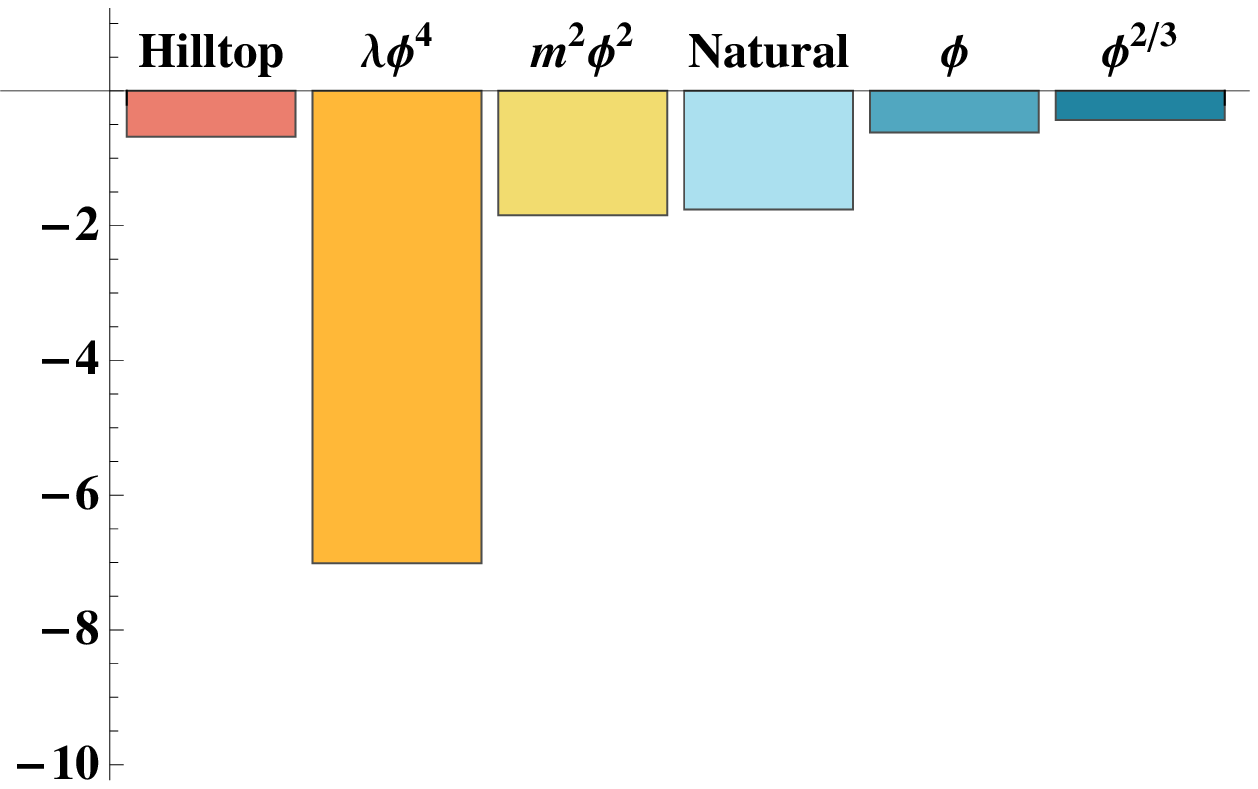, width=3in}}
 
 \mbox{}
 
\centerline{\epsfig{file=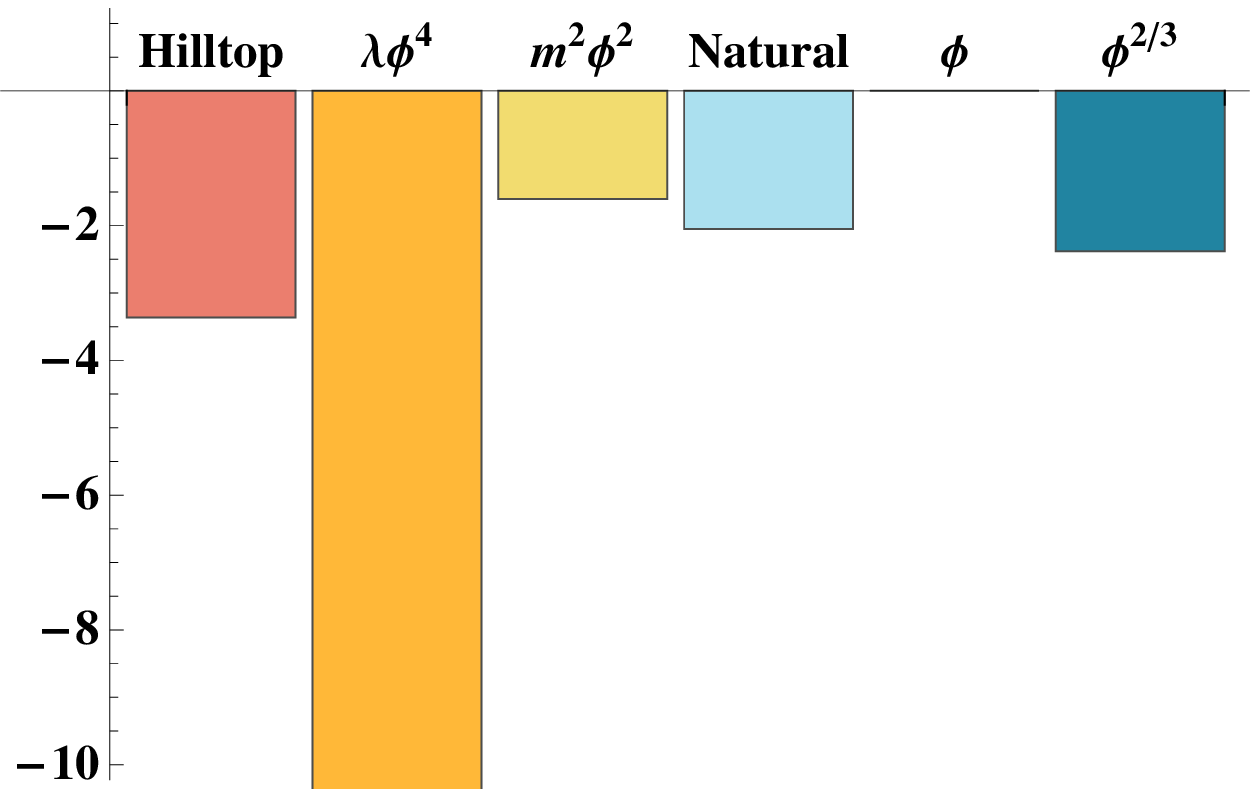, width=3in} \epsfig{file=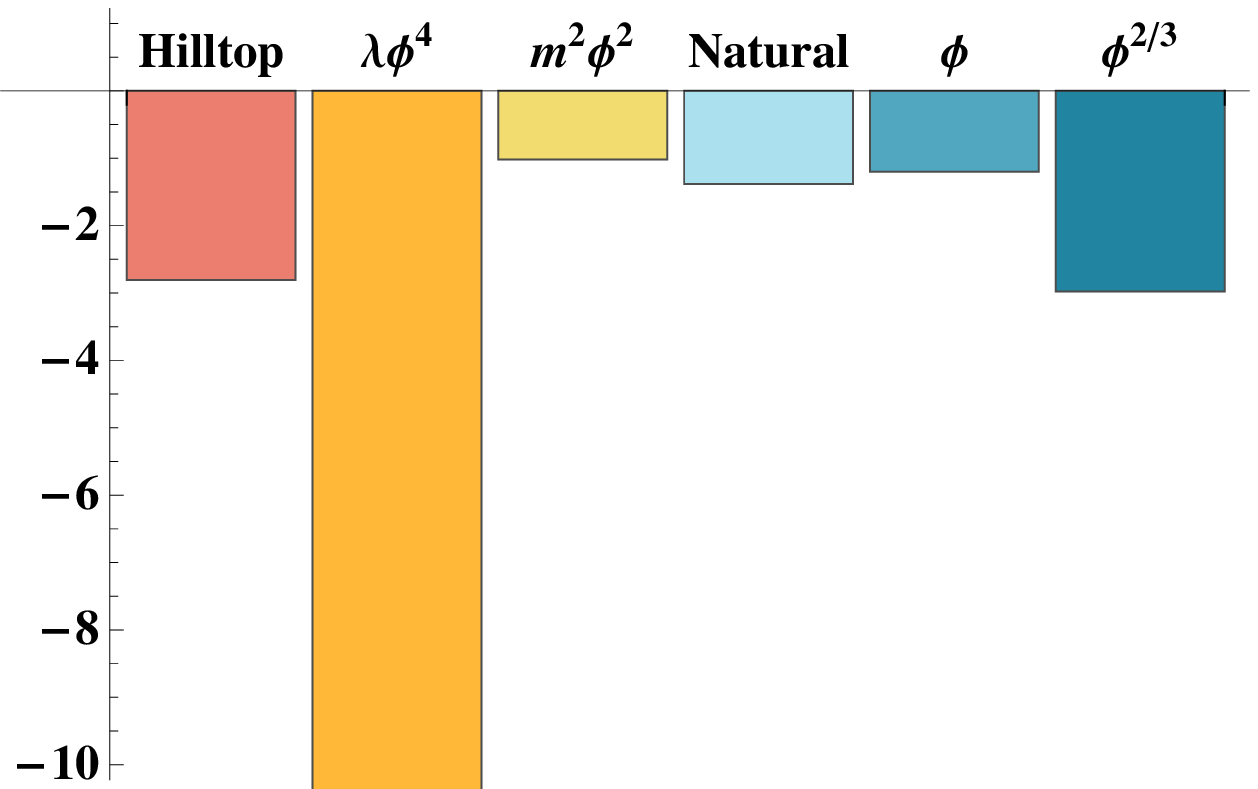, width=3in}}
\caption{ Bayesian evidence ($\Delta  \ln(E)$)  computed for the WMAP7 likelihood (top) and simulated Planck likelihood (bottom).  The left panels assume instant reheating. The right panels allow $-1/3 \le \wprim \le 1$ and $\rhorh^{1/4} \ge 100$ MeV.   $\Delta  \ln(E)$ is plotted relative to $V(\phi)\sim \phi$ inflation with instant reheating. The computed evidence values have a typical estimated uncertainty of $\sim 0.2$ for the \MultiNest\ settings used.  The charts are truncated at $\Delta  \ln(E)=-10$; with the Planck simulation we find that $\Delta  \ln(E)=-26.4$ and $-16.1$ for $\lambda \phi^4$ with instant  reheating and $\rhorh^{1/4} \ge 100$ MeV cases, respectively.   \label{fig:EV} }

\epsfig{file=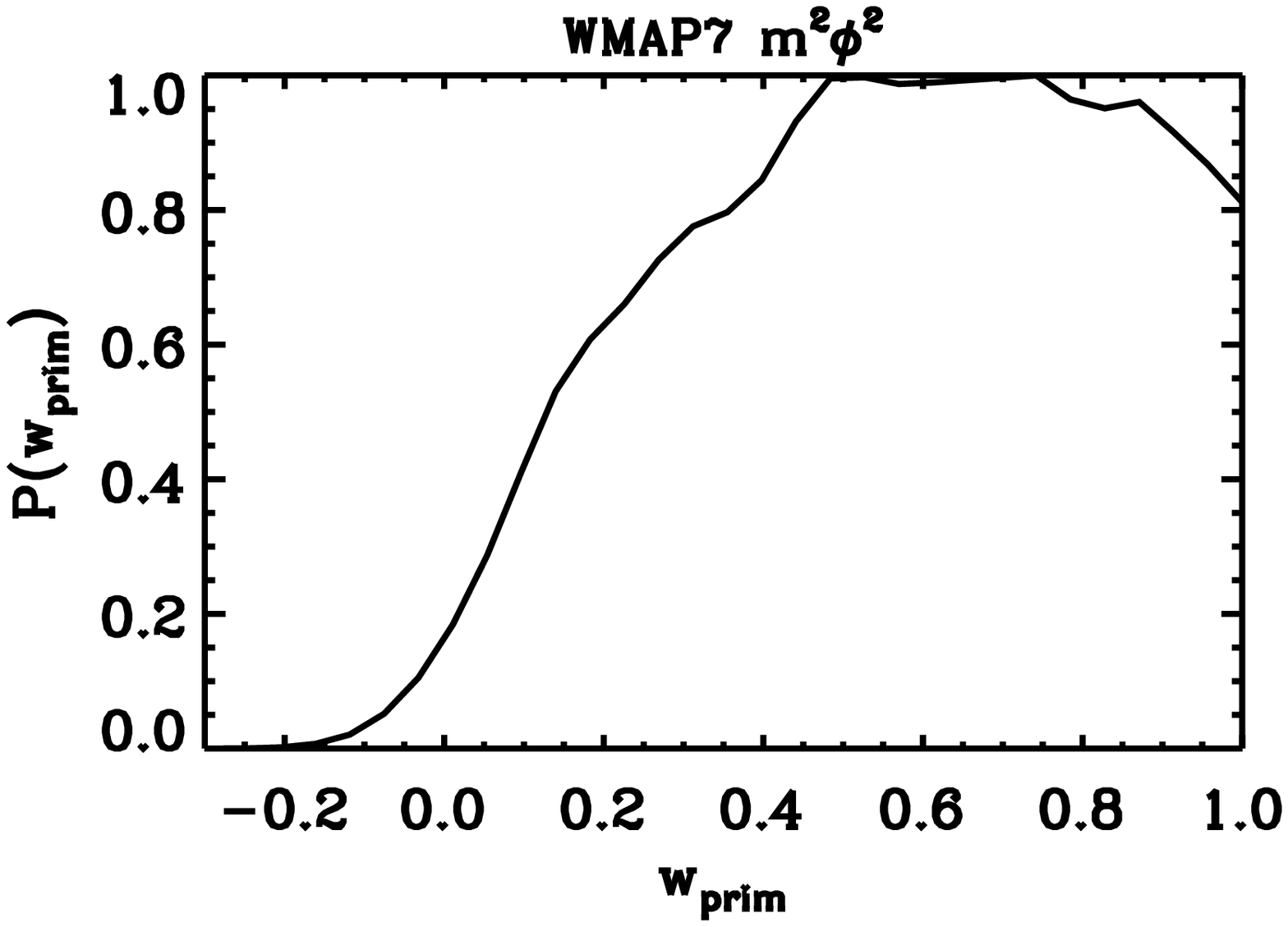, width=2.3in}
\epsfig{file=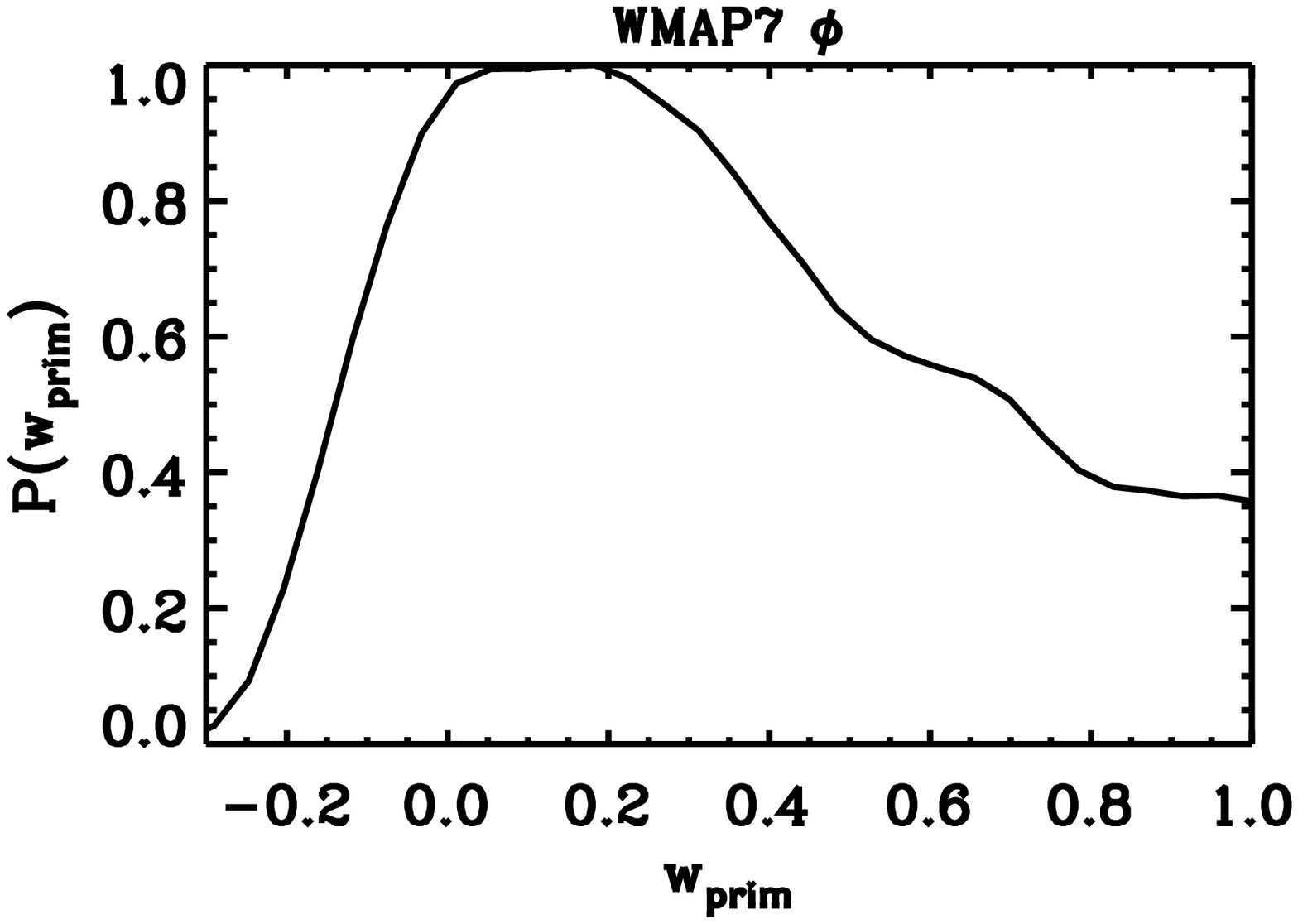, width=2.3in}
\epsfig{file=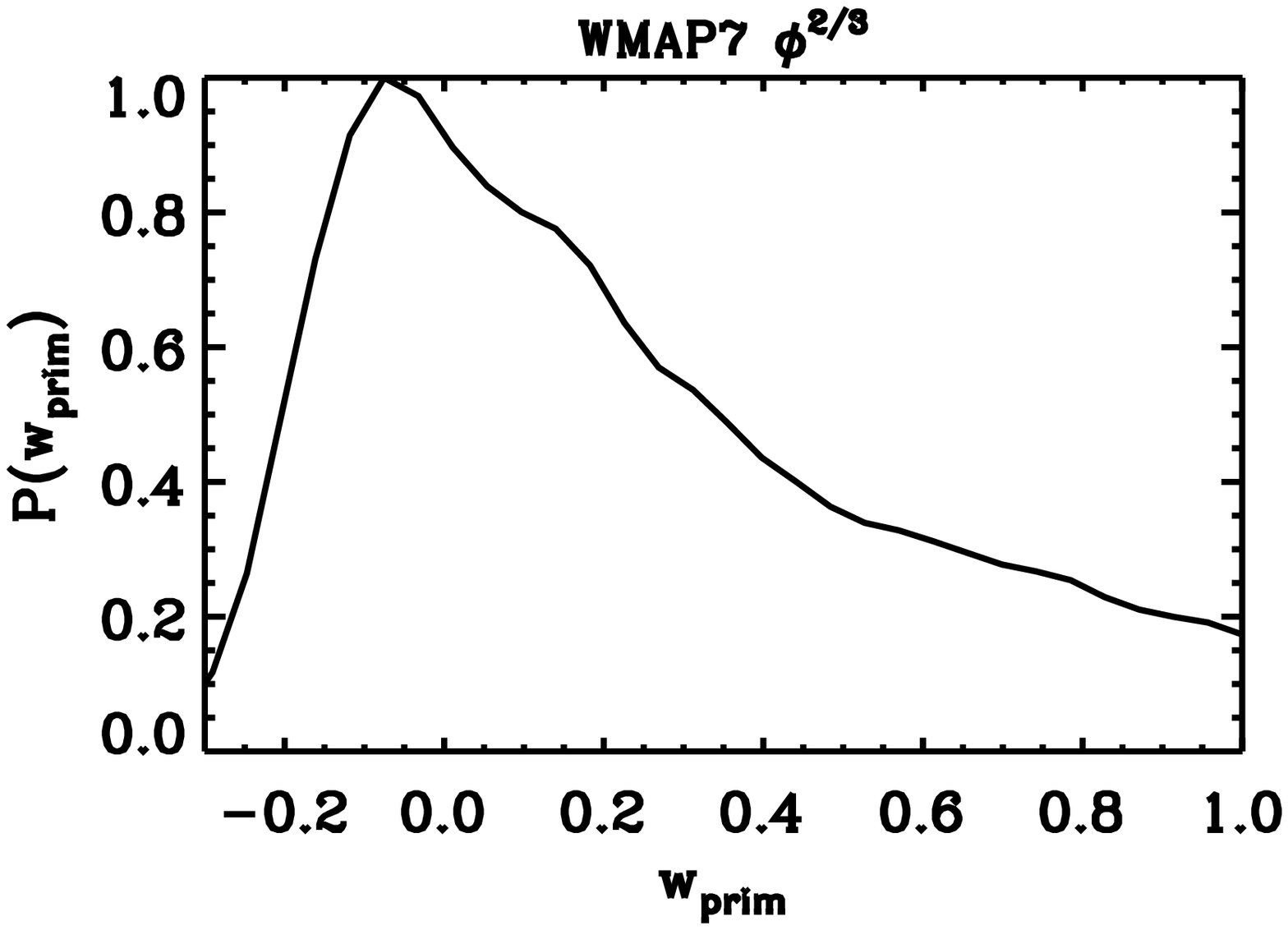, width=2.3in}

\mbox{}

 \epsfig{file=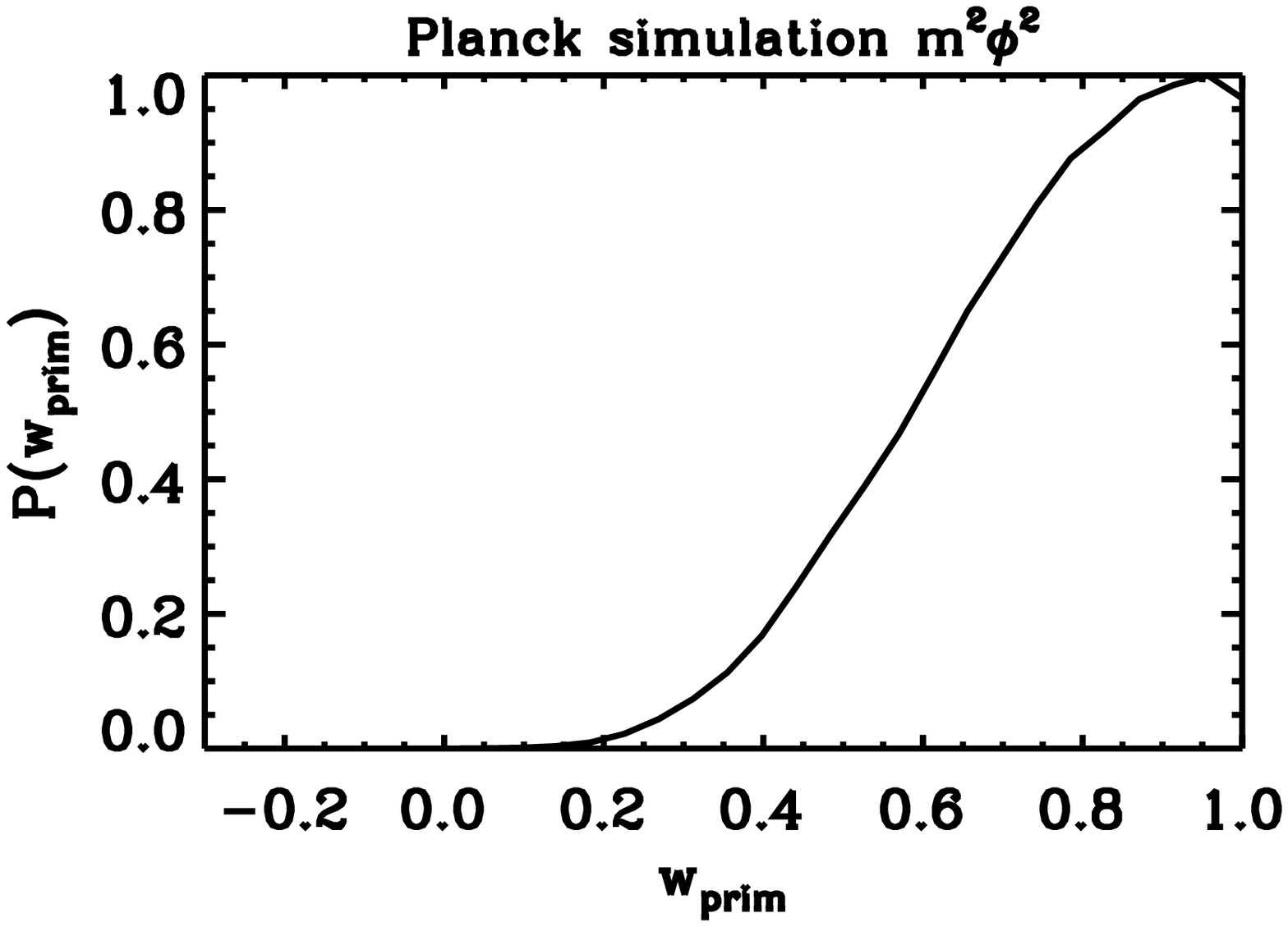, width=2.3in}
\epsfig{file=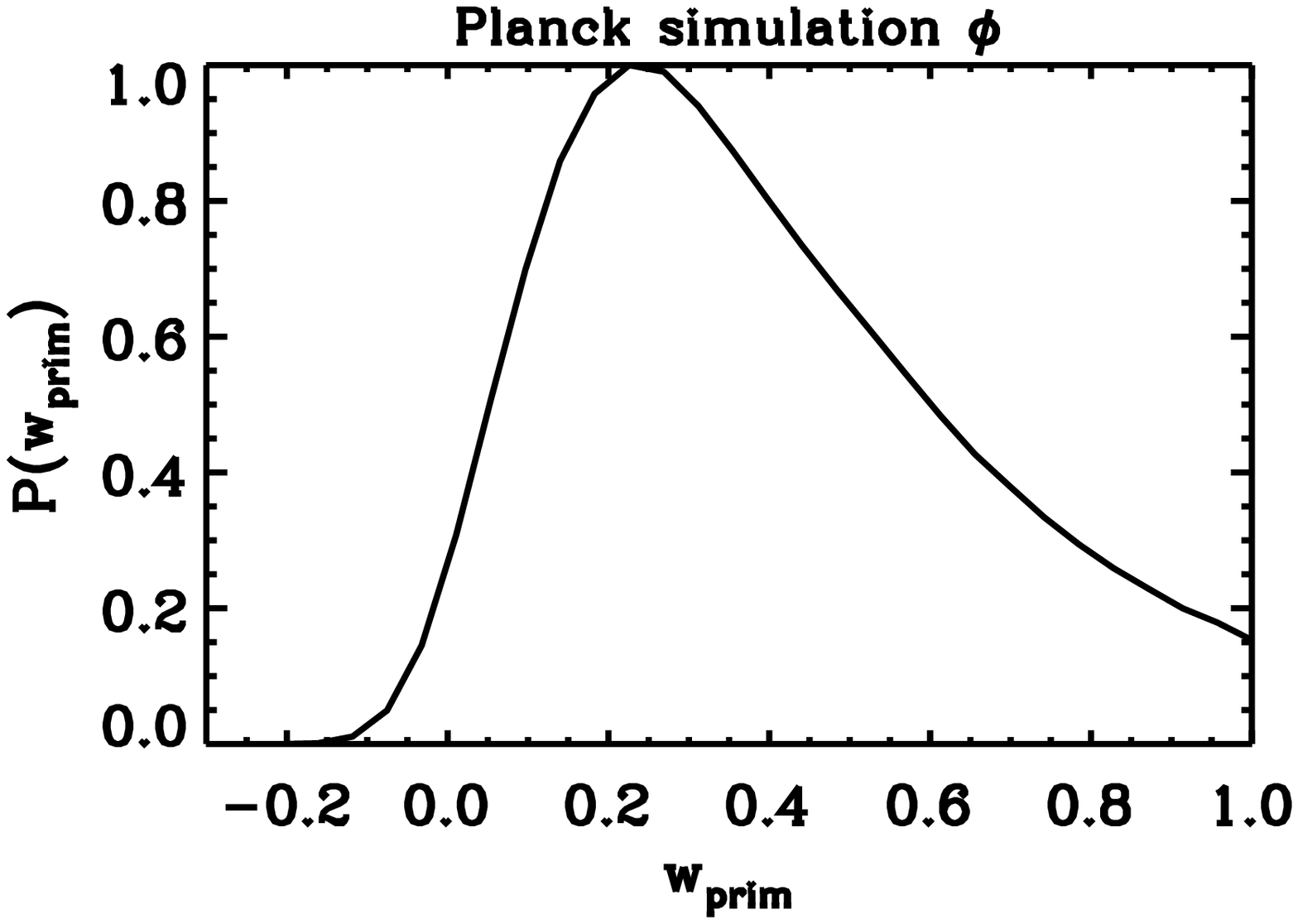, width=2.3in}
\epsfig{file=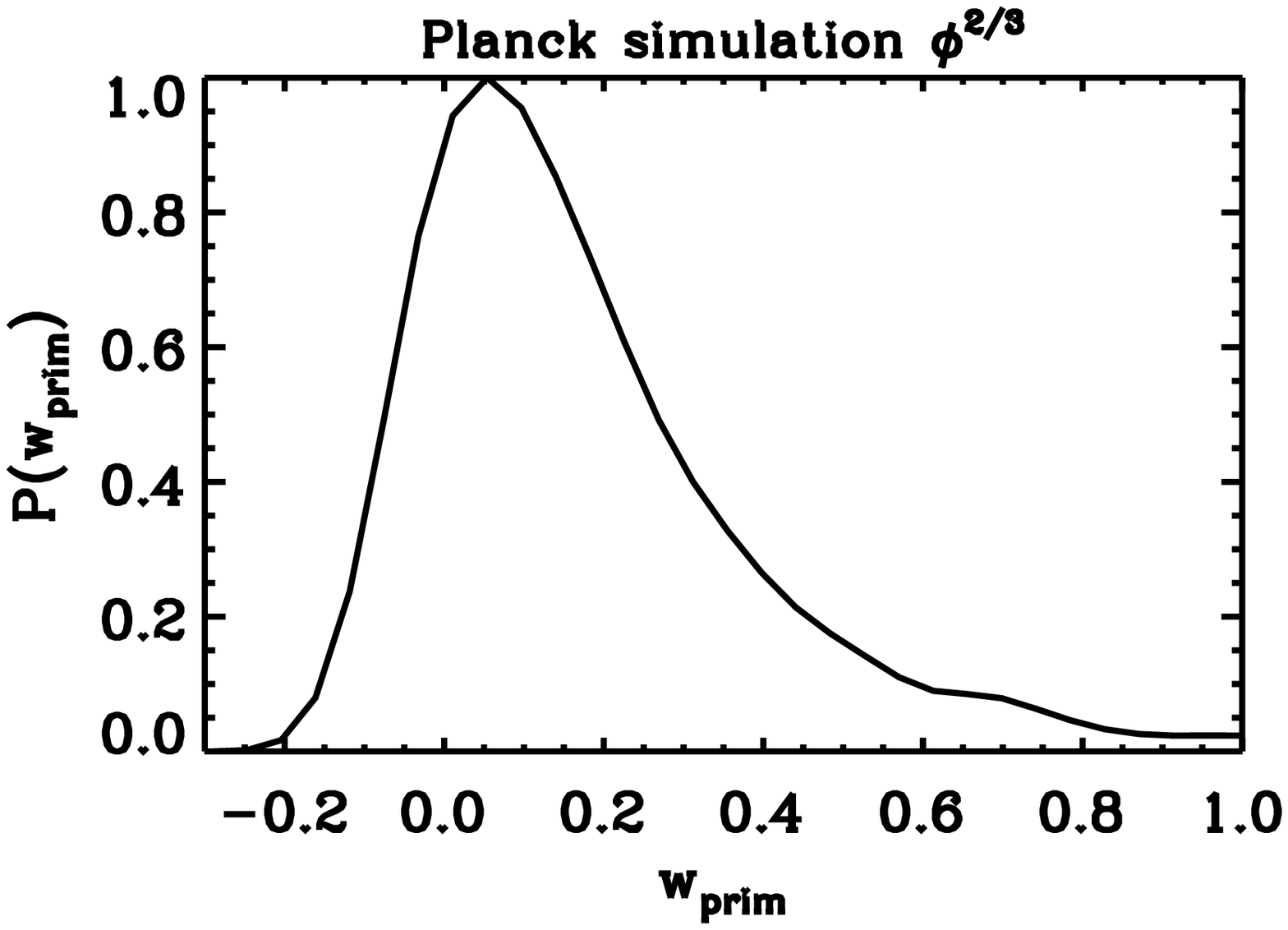, width=2.3in}
\caption{Posteriors for $\wprim$ from WMAP data (above) and the Planck simulation (below).   \label{fig:wplots} }

\end{figure*}

%

\section{Bayesian evidence and Parameter Constraints: Results and Forecasts}

We have computed Bayesian evidence for both 7 year WMAP dataset (WMAP7) \cite{Komatsu:2010fb},  and a simulated Planck likelihood, kindly provided by George Efstathiou and Steven Gratton (also utilized in Paper I). The Planck simulation assumes a primordial CMB power spectrum centered on the best fit WMAP5 cosmology (including $n_s=0.963$), with contributions from unresolved point sources and Sunyaev-ZelÕdovich (SZ) clusters, and instrumental noise, along with a tensor component with $r=0.1$.\footnote{Note that the specific realization of the sky generated for this simulation does not reproduce these central values exactly.} The Planck simulation is not used to make ``Fisher-style" forecasts, but instead analysed using the pipeline as if it were real data. This simulation is expected to broadly capture the constraining ability of the real Planck data, although the specific cosmology is, of course, a pre-launch guess.

\begin{figure*}[tb]
\centerline{\epsfig{file=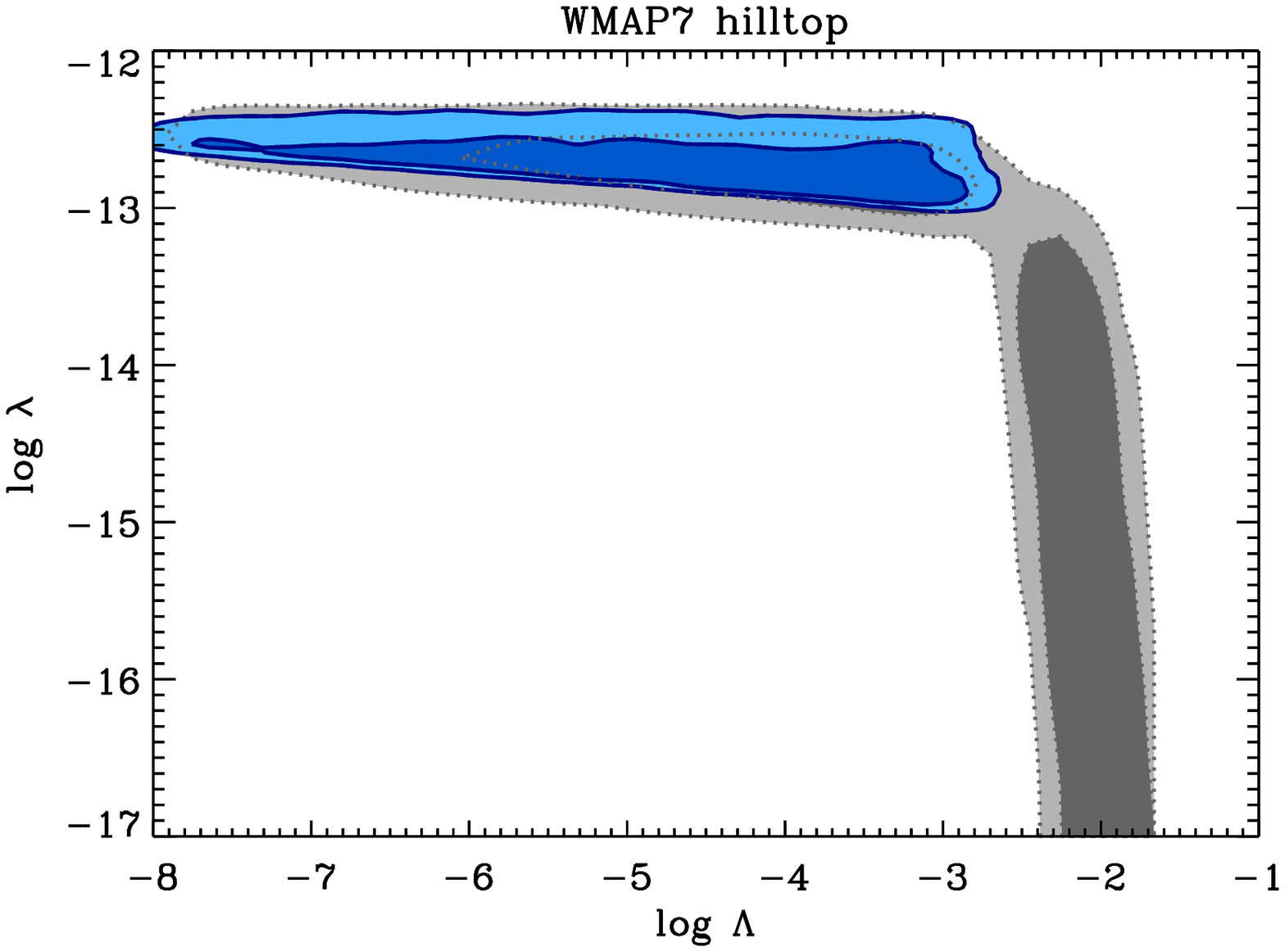, width=3.5in} 
\epsfig{file=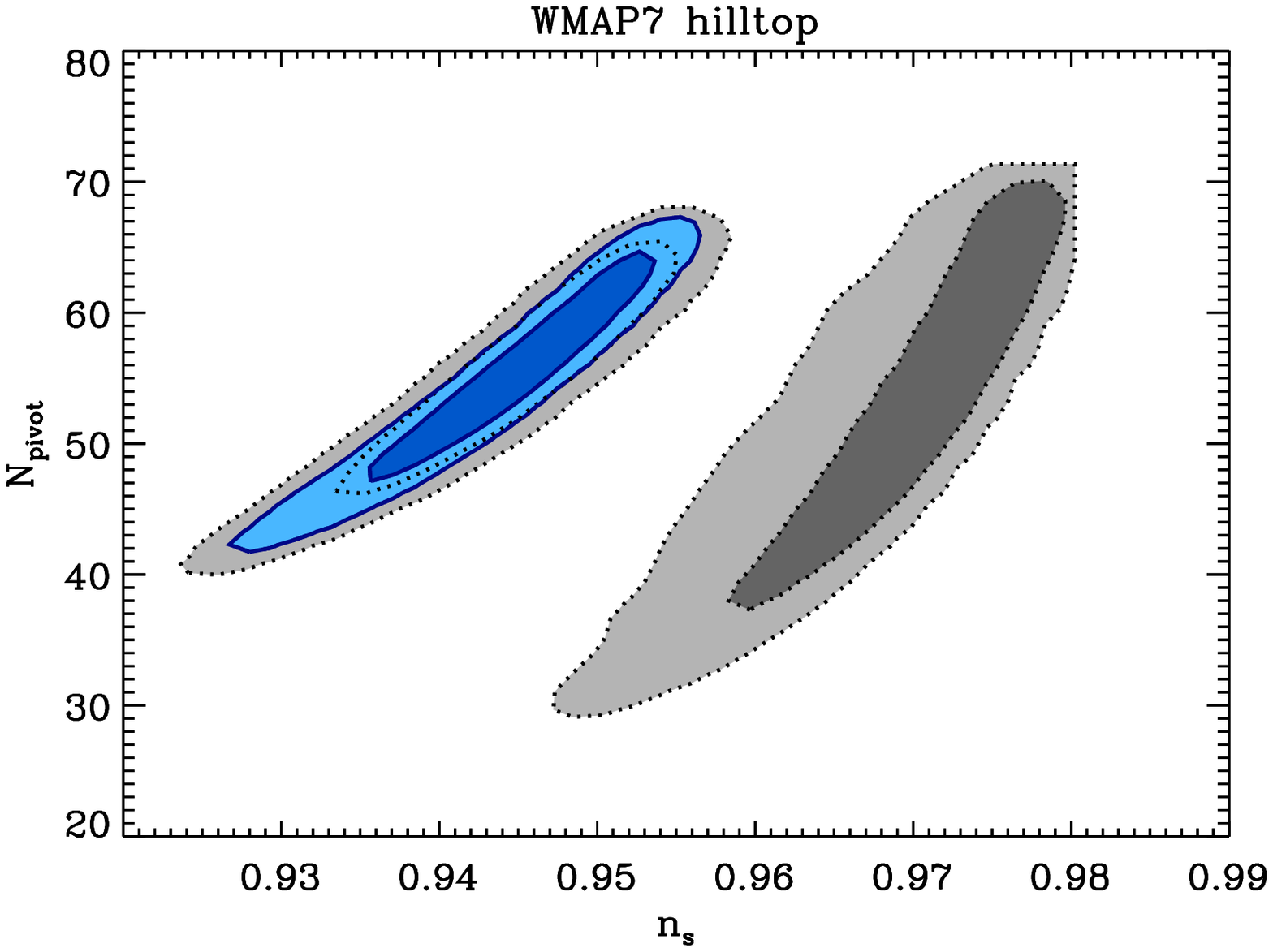, width=3.5in}}
\centerline{\epsfig{file=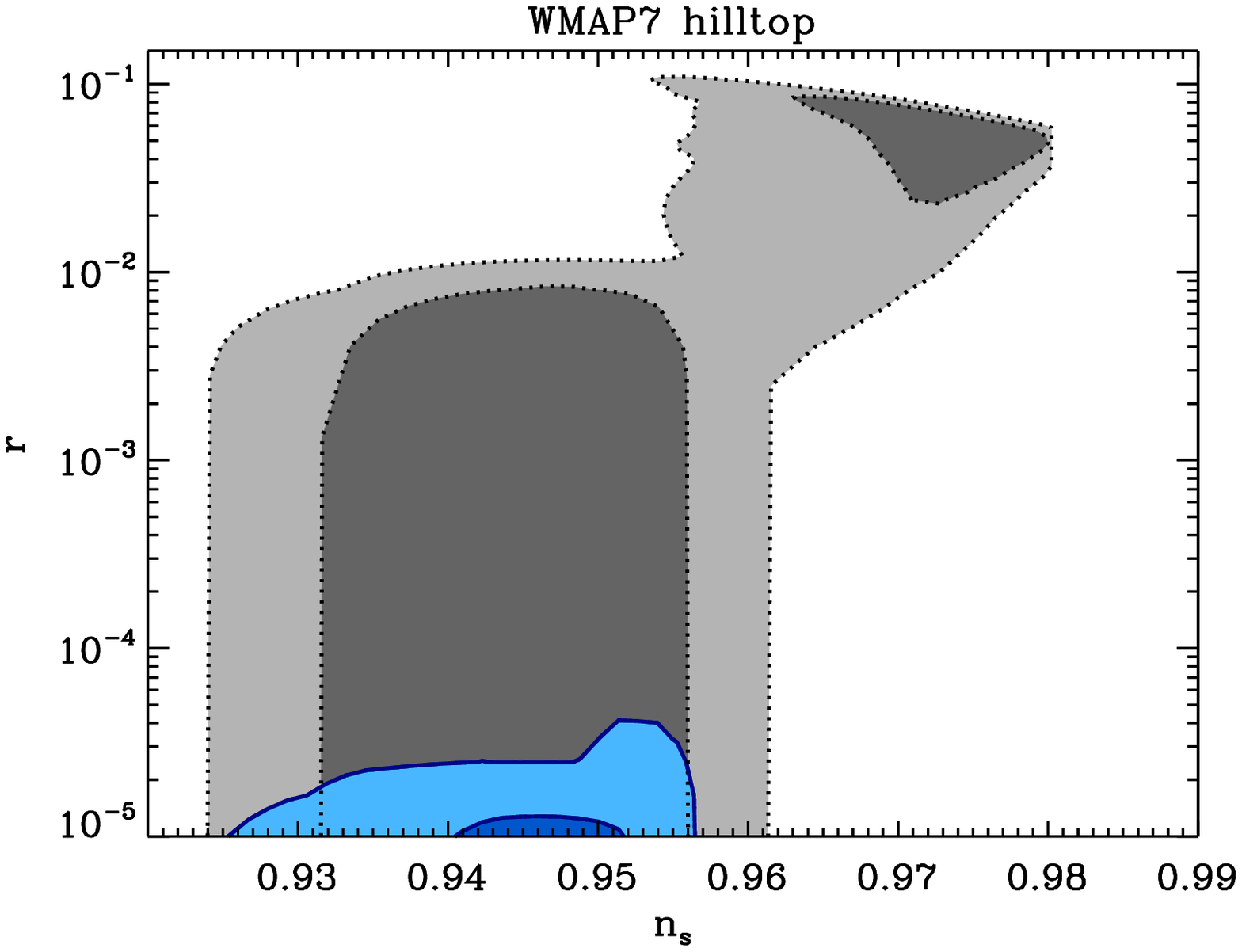, width=3.5in}
 \epsfig{file=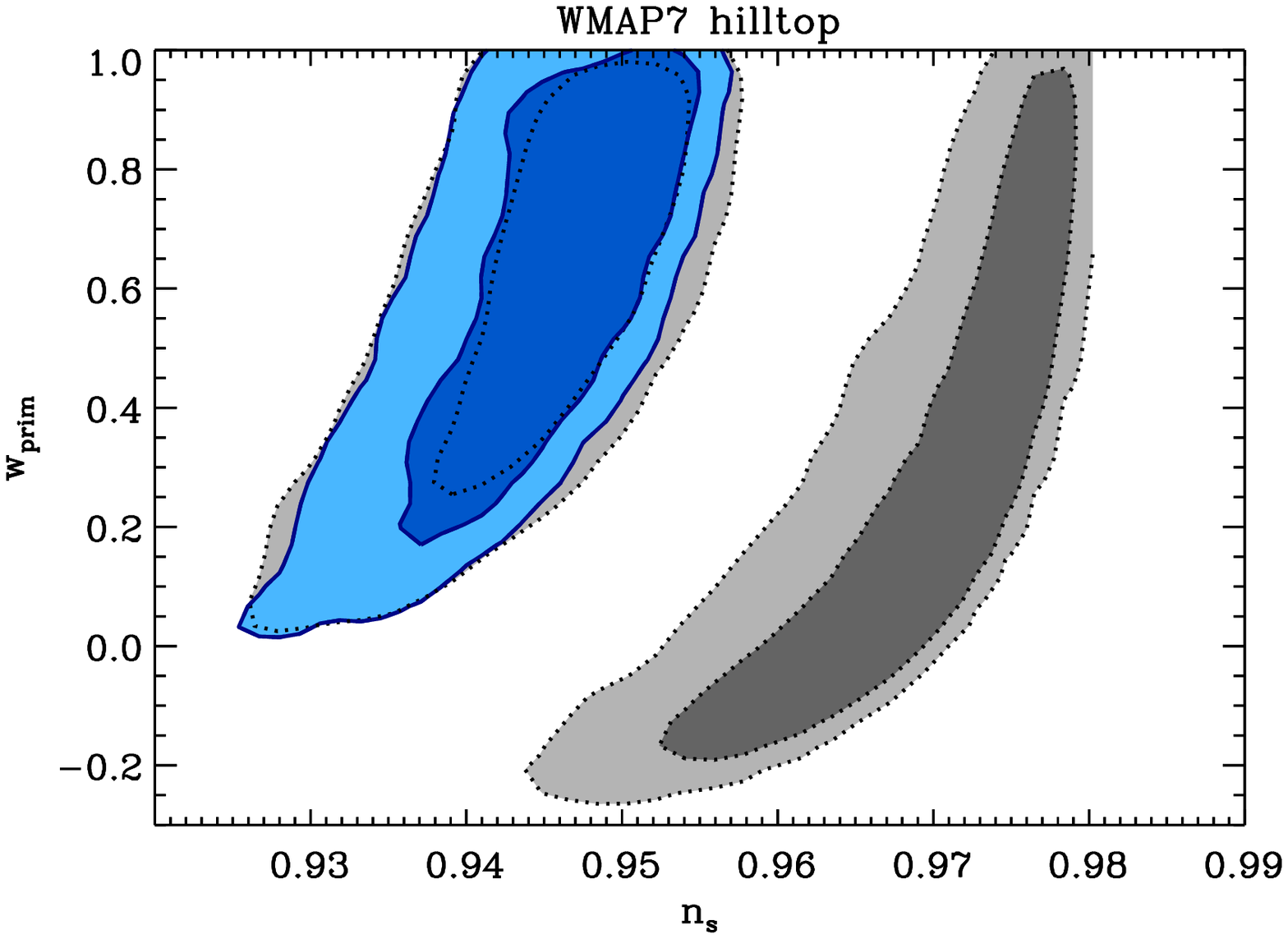, width=3.5in}}
\caption{ Parameter constraints obtained for hilltop inflation, with the WMAP7 likelihood. Blue regions denote the ``small field'' limit, while grey regions denote the posterior for the full hilltop prior from Table~\ref{tab:priors}. Contours correspond to $68$\% and $ 95$\% joint confidence levels. .  
\label{fig:hilltop}}
\end{figure*}

\begin{figure}[tb]
\centerline{\epsfig{file=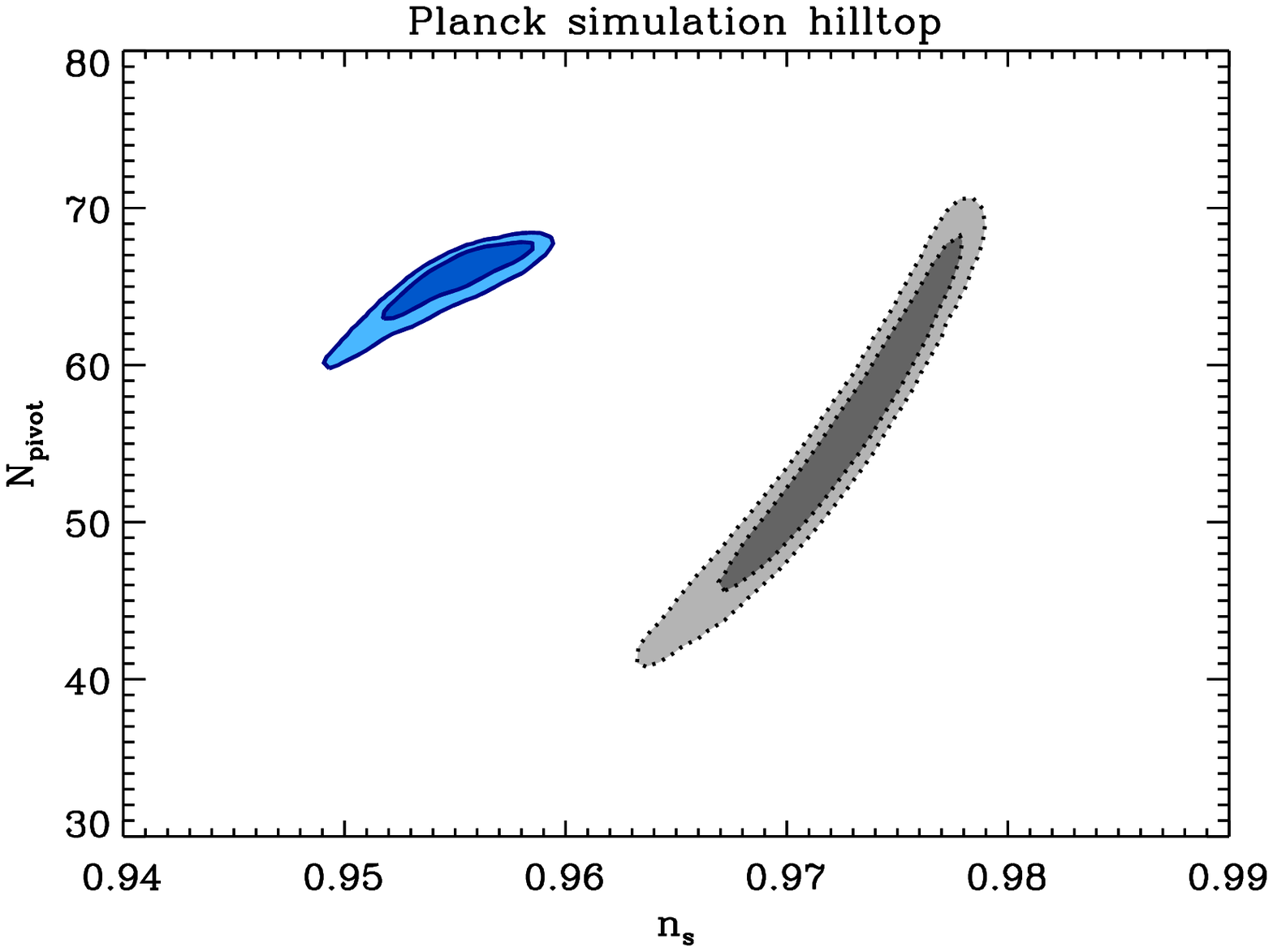, width=3.5in}}
\centerline{ \epsfig{file=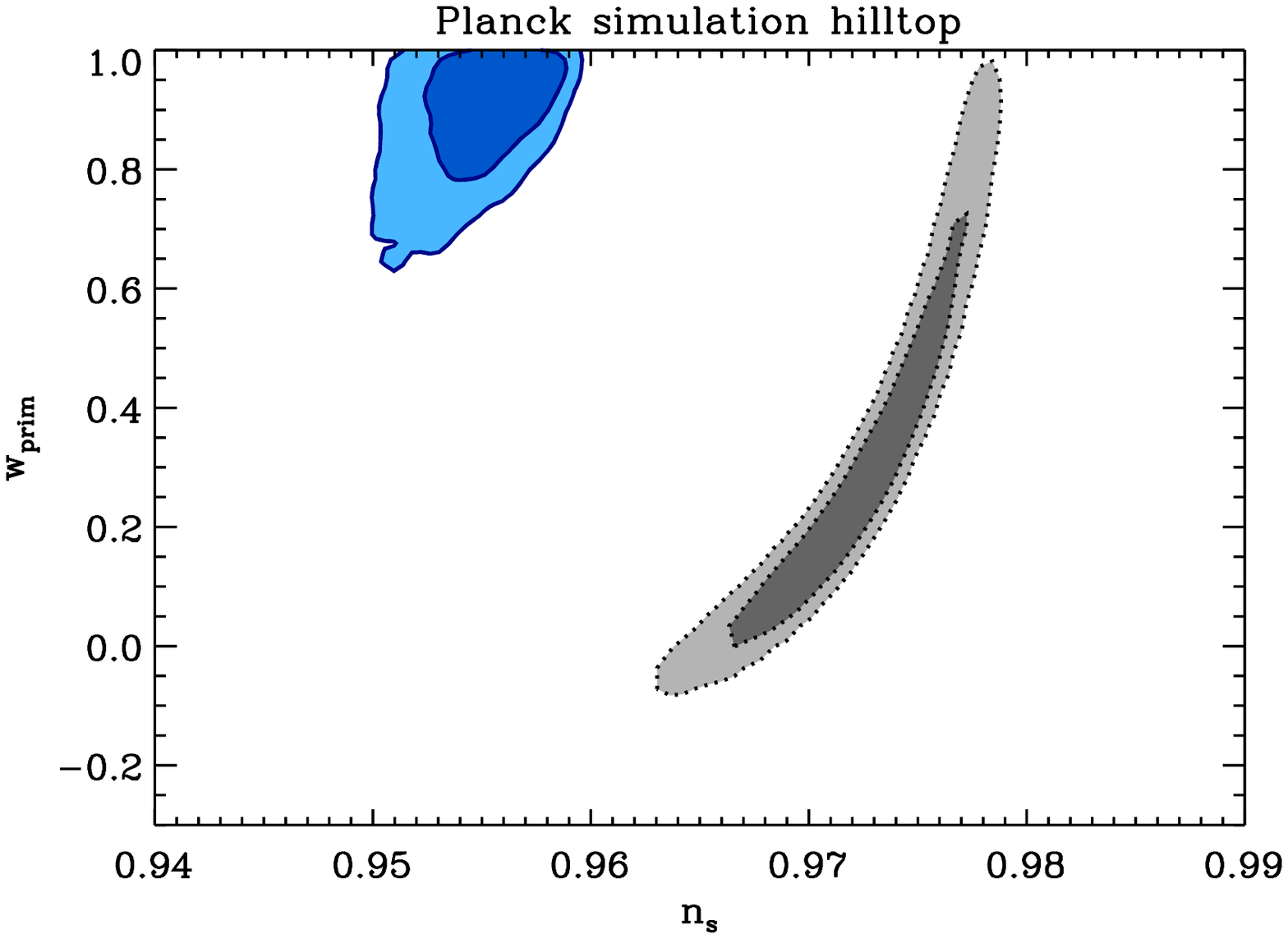, width=3.5in}}
\caption{ Parameter constraints obtained with the simulated Planck likelihood for hilltop inflation. Blue regions denote the ``small field'' limit, while grey denotes the posterior for the full hilltop prior from Table~\ref{tab:priors}. Contours correspond to $68$ and $ 95$\% joint confidence levels. In this case, the  small field region is strongly disfavored by the (simulated) data, and the posterior for the full hilltop prior is unimodal. In addition, in the small field limit $n_s$ becomes closer to the central value of the Planck simulation as $\Npiv$ increases, driving the posterior on $\wprim$ close to its maximal value.
\label{fig:hilltopplanck}}
\end{figure}

Figure~\ref{fig:EV} shows the evidence computed for the set of models we consider. Using the WMAP likelihood, only $\lambda \phi^4$ stands out as being strongly disfavoured; for the rest of our models  the  $\Delta  \ln(E)$ is typically only slightly greater than unity.\footnote{Note that these results are broadly consistent with those of Ref. \cite{Martin:2010hh}.}   Using the simulated Planck likelihood (Figure~\ref{fig:EV}) we see $\Delta  \ln(E) \sim 2$ for the same set of models, confirming our expectations that Planck will make nontrivial distinctions between models that are effectively degenerate when tested against present data.  Further, it should be pointed out that the Planck simulation does not directly match the predictions of any of the scenarios under consideration, so $\Delta  \ln(E) \sim 2$ is a conservative estimate of Planck's ability to discriminate between  models.

Parameter estimation for inflationary models using \ModeCode\ was the focus of Paper~I, and we will not repeat those results here. However, a significant addition to the current iteration of \ModeCode\ is the ability to compute $\wprim$,  and  estimate the posterior distribution for this parameter.   Figure~\ref{fig:wplots} show the constraints on $\wprim$ derived from WMAP7 and the Planck simulation, respectively.  With a quadratic potential the posterior for $\wprim$ is peaked in the region $\wprim>1/3$, and  this preference sharpens substantially with the simulated Planck likelihood. This provides further evidence that Planck will discriminate between different implementations of the same model of inflation -- these scenarios will necessarily have the same potential, $V(\phi)$ but can differ in their predictions for the post-inflationary universe.  For example,  a scenario where quadratic inflation was followed by moduli production and  a lengthy period of matter domination followed in turn by thermal inflation will have $-1/3 < \wprim < 0$.  Given the simulated  Planck likelihood used here, a universe with this post-inflationary history would be significantly disfavored relative to an implementation of quadratic inflation with prompt thermalization.

The evidence values we have computed for the Planck simulation suggest that, while the next generation of CMB experiments will provide substantial insight into the inflationary model selection problem, they will not uniquely single out a specific inflationary scenario as being compatible with the data. However, within  scenarios with two or more distinct inflationary regimes (e.g. the large and small field limits of hilltop inflation), Planck will often be able to eliminate one of these possibilities. 

 For example, the hilltop model has two distinct limits -- the ``small field'' scenario, where $r\lesssim 0.001$, and a ``large field'' limit in which $V(\phi) \sim \phi$. Our prior allows both cases. In the small field limit $n_s< 0.95$, and in the large field limit $V(\phi) \propto \phi$ \cite{Adshead:2010mc}. Both limits are compatible with the WMAP7 likelihood, but Planck would  exclude the small field limit if the real universe conforms to the parameters of the  simulation.  Physically, the small field limit is consistent with the Lyth bound, and we can select it by fixing $\log_{10}\Lambda< -2.5$ in the prior. For this choice $\Delta  \ln(E) \sim 40$ (relative to $V(\phi) \propto \phi$)  for the Planck simulation, from which we  infer  that the small field regime of the hilltop model is disfavored with very high confidence.

 Figure~\ref{fig:hilltop} shows posterior distributions obtained from \ModeCode\ for hilltop inflation with the WMAP7 likelihood.  This plot demonstrates the ability of the \MultiNest\ sampler to explore models where the likelihood contours are highly nontrivial: in Paper~I, the prior for the hilltop model was restricted  to the small field limit, so that the MCMC sampler converged in a finite amount of time.  
 
As implemented here, \ModeCode\  performs parameter estimation and computes  evidence in 10--20 hours on a single cluster node for a simple model -- approximately 3-4 times longer than required for a standard \CosmoMC-based parameter estimation (given a pre-computed covariance matrix) on the same hardware.  Estimates for two parameter models with Planck-quality data can require several node-days.  However,  MCMC chains become far less efficient when the parameter contours are topologically nontrivial: for natural inflation \MultiNest-based parameter estimation is substantially more efficient than with the MCMC chains presented in Paper I. Interestingly, using the WMAP likelihood, the posterior for $n_s$ is strongly bimodal --- this parameter changes quickly in the ``elbow'' region of the $(\log_{10}(\Lambda), \log_{10}(\lambda))$ plane (which is sampled with a uniform prior), leading to the strongly peaked posterior shown in Figure~\ref{fig:hilltop}. Conversely, the simulated Planck data effectively eliminates the small field branch of the hilltop parameter space, as seen in Figure~\ref{fig:hilltopplanck}.

\section{Discussion}  

In this Paper we have described the use of the \ModeCode\ solver for inflationary observables with the \MultiNest\ sampler, and the computation of Bayesian evidence for inflationary models. For a representative sample of single field models,  evidence computed using the WMAP7 likelihood only permits a definitive conclusion regarding $\lambda \phi^4$ inflation -- reproducing a well-known result --  while the $\Delta \ln(E)$ obtained for other models were not large enough discriminate between these models even tentatively.

Our primary motivation here is to survey the issues associated with specifying physically realistic priors for inflationary models in advance of the Planck dataset becoming available.  Using evidence computed with a likelihood derived from a simulated Planck dataset we have shown that Planck will  indeed begin to discriminate between models which are effectively  degenerate from the perspective of WMAP.  However, great care must be taken when specifying the priors, in order to ensure that these correspond to statements which are a) physically realistic and b) genuinely independent of the data used to construct the likelihood, $\mathcal{L}$.  In particular, priors (and notably, the allowed ranges of the free parameters) which are acceptable when performing {\em parameter estimation\/} are not physically justifiable when computing evidence. Unrealistic  prior ranges can bias the computed values of the evidence, as well as dramatically reducing the computational efficiency of \MultiNest.  Furthermore, we have proposed  an approach to specifying  priors for multi-parameter inflationary models which simultaneously excludes models with grossly unphysical perturbation spectra, while retaining a strictly uniform prior upon the inflationary parameters themselves.  

This is not the first discussion of cosmological model selection via Bayesian evidence and, in particular, this topic was recently  treated in Ref. \cite{Martin:2010hh}. However, the present analysis breaks new ground in several key directions. We highlight the importance of choosing physically reasonable priors -- both to produce self-consistent results, and  to ensure the code is numerically efficient. In addition we provide an algorithm capable of  computing evidence for general multi-parameter potentials, without which we would be unable to compute evidence for the natural or hilltop cases. Lastly, since theory does not specify precise ranges for the parameters in the models, we propose a method which provides self-consistent, maximum entropy priors, ensuring that all models are compared on an equal footing. 

Planck is expected to provide an exquisitely accurate measurement of $n_s$  but, given the values of $\Delta \ln(E)$ computed for the Planck simulation,  it will not necessarily resolve the inflationary model selection problem. This is true even for inflationary scenarios that predict values of $n_s$ which  differ substantially from one other, relative to precision with which Planck is expected to measure this quantity.  However, none of the scenarios  here  yield values of $n_s$ and $r$ which fully overlap  with the central values  in the Planck simulation.    

In principle, additional data will break any degeneracy in the evidence values for a set of inflationary models.  Planck-based constraints of inflationary models are primarily due to the exquisite measurement of the scalar perturbations  that this satellite is expected to provide. Conversely, constraints on the primordial tensor (gravitational wave) amplitude $r$ are likely to be significantly tightened by several forthcoming balloon and ground-based experiments. The models evaluated here  generate a range of values of $r$, and a joint analysis of the Planck dataset and future polarization-sensitive experiments should  significantly increase  the range of evidence values, relative to those computed here.

It is clear from our simulations that Planck will put strong and nontrivial constraints on the parameter values of specific inflationary scenarios. Thus for models (natural and hilltop inflation, in our examples) with two observably distinct inflationary regimes, Planck is likely to  discriminate between these possibilities. Further, the predictions of  inflationary models necessarily depend on the assumed post-inflationary expansion history  \cite{Adshead:2010mc,Mortonson:2010er}, which we characterize using the effective equation of state $\wprim$. The posterior distributions obtained for $\wprim$ from the simulated  likelihood confirms that Planck should be able provide interesting and nontrivial constraints on the post-inflationary expansion history.

\section*{Acknowledgments}
We thank Michael Mortonson for co-developing \ModeCode\ in Paper I, and Michael Bridges for advice on  \MultiNest. We are very grateful to George Efstathiou and Steven Gratton for access to their unpublished pre-launch Planck simulation, which was used to generate the Planck forecasts.  RE is partially supported by the United States Department of Energy (DE-FG02-92ER-40704) and the National Science Foundation (CAREER-PHY-0747868).  HVP is supported in part by Marie Curie grant MIRG-CT-2007-203314 from the European Commission, and by STFC and the Leverhulme Trust. RE thanks University College London for  hospitality while this paper was completed, and HVP and RE acknowledge the hospitality of the Benasque Science Center where part of the work was carried out. We acknowledge the use of the Legacy Archive for Microwave Background Data (LAMBDA). Support for LAMBDA is provided by the NASA Office of Space Science.

\mbox{}

\bibpreamble{\vspace{-1cm}}

\bibliographystyle{h-physrev3}
\bibliography{modecode.bib}


\end{document}